\documentclass[Afour,sagev,times]{sagej}

\usepackage{moreverb,url}
\usepackage{enumitem}
\usepackage{xspace}
\usepackage{nicefrac}
\usepackage[colorlinks,bookmarksopen,bookmarksnumbered,citecolor=red,urlcolor=red]{hyperref}
\usepackage{subcaption}
\usepackage{soul}
\usepackage{float}
\usepackage{graphicx}

\newcommand\BibTeX{{\rmfamily B\kern-.05em \textsc{i\kern-.025em b}\kern-.08em
T\kern-.1667em\lower.7ex\hbox{E}\kern-.125emX}}

\usepackage{tikz}
\DeclareRobustCommand{\ballnumber}[1]{\tikz[baseline=(myanchor.base)] \node[circle,fill=.,inner sep=1pt] (myanchor) {\color{-.}\bfseries\footnotesize #1};}

\newcommand{\ie}{{i.e.,}\xspace}
\newcommand{\etal}{{\em et al.}\xspace}

\newif\ifdraft
\drafttrue %for draft version
\usepackage{todonotes}      % annotations  

\def\volumeyear{2023}

\begin{document}

\title{Adversarial Attacks on Reinforcement Learning Agents for Command and Control}

\author{Ahaan Dabholkar\affilnum{1}, James Z. Hare\affilnum{2}, Mark Mittrick\affilnum{2}, John Richardson\affilnum{2}, Nicholas Waytowich\affilnum{2}, Priya Narayanan\affilnum{2} and Saurabh Bagchi\affilnum{1}}

\affiliation{\affilnum{1} Purdue University\\
\affilnum{2} DEVCOM Army Research Laboratory, USA}

\corrauth{Ahaan Dabholkar, Purdue University}

\email{adabholk@purdue.edu}

\begin{abstract}
Given the recent impact of Deep Reinforcement Learning %agents 
in training agents to win complex games like StarCraft and DoTA~(Defense Of The Ancients) -- there has been a surge in research for exploiting learning based techniques for professional wargaming, battlefield simulation and modelling. Real time strategy games and simulators have become a valuable resource for operational planning and military research. However, recent work has shown that such learning based approaches are highly susceptible to adversarial perturbations. In this paper, we investigate the robustness of an agent trained for a Command and Control task in an environment that is controlled by an active adversary. The C2 agent is trained on custom StarCraft II maps using the state of the art RL algorithms -- A3C and PPO. We empirically show that an agent trained using these algorithms is highly susceptible to noise injected by the adversary and investigate the effects these \textit{perturbations} have on the performance of the trained agent. Our work highlights the urgent need to develop more robust training algorithms especially for critical arenas like the battlefield.
\end{abstract}

\keywords{Adversarial Attacks, Deep Reinforcement Learning, Command and Control, Starcraft II, Adversarial Robustness}

\maketitle

 \section{Introduction}

Deep Reinforcement Learning~(DRL) has been successfully used to train agents in several tactical and real-time strategy games such as StarCraft\cite{starcraft} and DoTA\cite{dota}, which involve complex planning and decision-making. These agents have demonstrated proficiency in coming up with winning strategies comparable to that of experienced human players (AlphaStar~\cite{vinyals}, OpenAI Five\cite{openai2019dota}) through techniques like self--play, imitation learning, etc. 
As a result, in recent years, there has been mounting interest in the military research community in applying these RL techniques to tasks such as operational planning and command and control~(C2). Simultaneously, traditional game engines have been repurposed to facilitate automated learning (pySC2\cite{pysc2}, SMAC\cite{samvelyan2019starcraft, smacgithub}, pyDoTA\cite{pydota2}) and new ones developed for battlefield simulation~\cite{narayanan2021first,park2022deep,soleyman2020multi,zhang2020air,BZCFHSWA2022}, 
creating what are effectively \textit{digital wargames}. The driving force behind this research has been to improve and augment strategies used on the battlefields of the future, which are expected to be more complex and unconventional, possibly beyond the cognitive abilities of a human commander.

Recent works~\cite{waytowich2022learning} have had considerable success in winning simulated wargames using C2 agents that have been trained through reinforcement learning techniques and synthetic data. This has been possible partly due to the scalability of RL training which has proved to be a massive advantage for exploring and exploiting different strategies, even when faced with difficult or complicated scenarios and given only partial information about the environment. 
However, these evaluations are done in benign environments where information available to the C2 agent is assumed to be {\em uncorrupted}. Realistically, this is unlikely in situations on the battlefield where information may have inherent noise because of the mode of collection~(from sensors or other sources) or may be tampered with by the enemy. In this work, we evaluate the robustness of such a trained agent when subject to potentially adversarial inputs in the context of C2.

\begin{figure}[!t]
    \centering
    \includegraphics[width=1.0\linewidth]{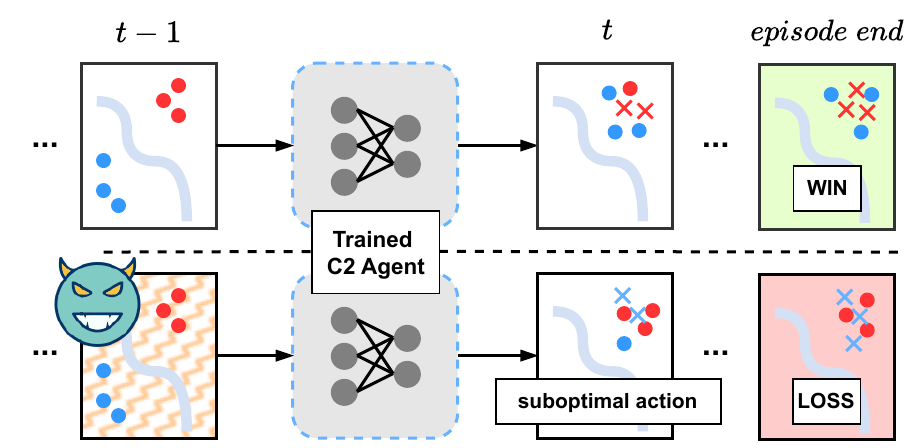}
    \caption{\textbf{Robustness Evaluation Methodology:} The figure shows the difference between a benign \textit{\textbf{(top)}} and malicious \textit{\textbf{(bottom)}} environment at timestep $t$. Observations at $t-1$ are input to a C2 agent that has been pretrained in a benign environment. The agent samples a suboptimal action as a result of the injected adversarial perturbations (orange) in the input which eventually leads to a loss for the BlueForce.}
    \label{fig:intro-fig}
\end{figure}

In order to do so, first, we use the StarCraft II Learning Environment~(SCLE\cite{vinyals2017starcraft}) to model conflict between two teams, the BlueForce and the RedForce. The C2 agent directs the BlueForce to win the battle by eliminating the RedForce troops. 

Next, we assume the an attacker present in the environment tampers with the observations collected from the battlefield before they are made available to the C2 agent. The added perturbations termed \textit{adversarial perturbations} are constructed to be highly imperceptible to evade detection while maximally subverting the C2 agent's policy to something deleterious~(Figure~\ref{fig:intro-fig}). We then evaluate the resulting drop in performance of the agent on several metrics as well as analyze the deviation in the course of action from a military perspective.

Our key contributions are summarized as follows:
\begin{itemize}[wide, labelindent=0pt, itemsep=0pt]
    \item We empirically show the vulnerability of the trained C2 agent to even small adversarial perturbations in the input observations. Our studies quantify some expected trends and bring out some non-obvious trends. For example, our studies reveal that partially trained agents appear to be more resistant to noise than fully trained agents.
    \item For generality, we evaluate the effectiveness of the attacks on two different scenarios which correspond to the C2 agent's task to attack and defend respectively.
    \item We also evaluate agents trained using two state-of-the-art RL algorithms, A3C and PPO, and comment on their robustness to injected noise.
    \item We provide interpretability to the model's outputs by profiling the shift in action distributions predicted by the policy network caused by the attacker's perturbations.
    
\end{itemize}

Our evaluations demonstrate the susceptibility of vanilla RL training algorithms to adversarial perturbations and the need for robust training mechanisms and sophisticated detection and prevention techniques especially for such critical scenarios.

The structure of the paper is as follows. First we provide brief backgrounds on the use of RL for C2 followed by a description of the StarCraft Environment and two custom scenarios -- TigerClaw and NTC in Section~\ref{sec:Rl-C2}, which we use for training our agent. 
In Section~\ref{sec:technical}, we describe the state and action space of our custom scenarios and the details of the RL agent. Section~\ref{sec:inf-time-attack} and Section~\ref{sec:eval} contain the attack methodology and the evaluations respectively. Finally we include a discussion on the need for utilizing adversarially robust training techniques and directions for future work.
\section{Background}
\subsection{RL for Command and Control (C2)}
\label{sec:Rl-C2}
Mission success in military C2 requires disseminating control actions to designated forces based on real-time Intelligence, Surveillance, and Reconnaissance (ISR) information of the operational environment, such as kinetic and non-kinetic (e.g., weather, political, economic, cultural) variables, and terrain information. Before the battle begins, the commander and their specialized staffing officers must develop a detailed mission plan encapsulated as a Course Of Action (COA). The development of a COA requires a detailed analysis of the operational environment, predictions of the opposing force’s COA (strategy), and wargaming to identify a friendly force COA that is finely tuned to meet the mission requirements~\cite{marr2001military}. Typical military planning is solely based on the commander and staffing officers, and their allocated time before the battle dictates the number of possible COAs that can be considered. Additionally, each potential COA must be wargamed and fine-tuned against a small set of opposing force COAs to identify strengths and weaknesses. However, this can result in suboptimal (heuristic-based) COA~\cite{s2000}. 

To circumvent this limitation, future military planning is envisioned to incorporate an Artificial Intelligent (AI) commander’s assistant that can generate and recommend COAs to aid in the military planning process. Recent developments in deep RL for strategy games provides a promising direction to develop control policies for C2~\cite{vinyals}. The algorithms developed allow an AI agent to learn the best control \emph{policy} that optimizes a predefined reward function by playing millions of simulated games through the exploration of many environmental state and action pairs. To extend these formulations for military C2 requires modeling, simulating, and wargaming a large number of battles faster than real-time in a virtual environment that emulates realistic combat characteristics. Furthermore, the existing RL algorithms must be adapted to handle a large number of heterogeneous actors, doctrine based control strategies, complex state and action space, and uncertain information. 

Previous work on RL for C2 found that the StarCraft II gaming engine, developed by Blizzard Entertainment~\cite{starcraft}, provides a simulation environment that can be militarized and used for prototyping an AI commander’s assistant~\cite{goecks2023games,waytowich2022learning}. The following subsections provide details of the StarCraft II environment and the scenarios considered in this paper.

\subsection{StarCraft II C2 Environment}
\label{sec:sc2env}
StarCraft II is a multi-agent real-time strategy game developed by Blizzard Entertainment~\cite{starcraft} that consists of multiple players competing for influence and resources with the ultimate goal of defeating the other players. As a collaboration between Deepmind and Blizzard Entertainment, the \emph{StarCraft II Learning Environment} (SC2LE) Machine Learning API was developed to allow researcher to study many difficult challenges associated with RL~\cite{pysc2}. For example, controlling heterogeneous assets in a complex state and action space with uncertainties. Furthermore, the \emph{StarCraft II Editor} allows developers to construct custom scenarios, making it possible to develop RL agents for C2. 

Previous work extended this framework to militarize the SC2LE and develop a C2 simulation and experimentation capability that interfaces with deep RL algorithms via RLlib~\cite{goecks2023games}, an open-source industry-standard RL library~\cite{liang2018rllib}. The icons were re-skinned to portray standard military symbology, new StarCraft II maps were designed to emulate realistic combat characteristics, such as environment terrain and asset attributes (e.g., visibility, weapons, weapons ranges, and damage), and a custom wrapper was designed to train RL agents. 
For a detailed description of this framework, see~\cite{goecks2023games}. 
In our work, we use this framework to develop baseline RL agents for C2 and study the effects of adversarial attacks on the learned policies in the following two scenarios.

\begin{figure}[!ht]
    \centering
    \fbox{\includegraphics[width=0.8\linewidth]{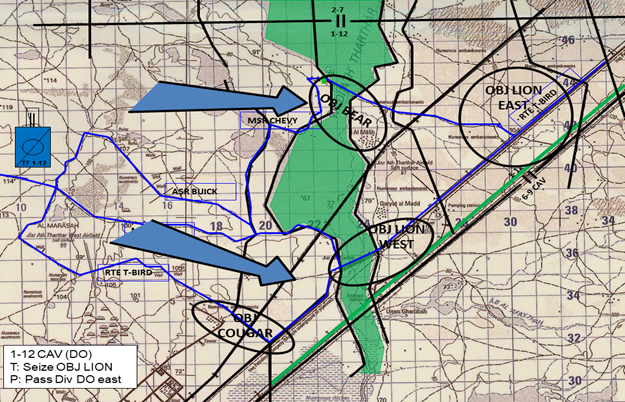}}
    \caption{TigerClaw Scenario}
    \label{fig:TigerClaw-Scenario}
\end{figure}

\subsubsection{Custom StarCraft II Scenarios}

\begin{figure}[!ht]
    \centering
    \fbox{\includegraphics[width=0.8\linewidth]{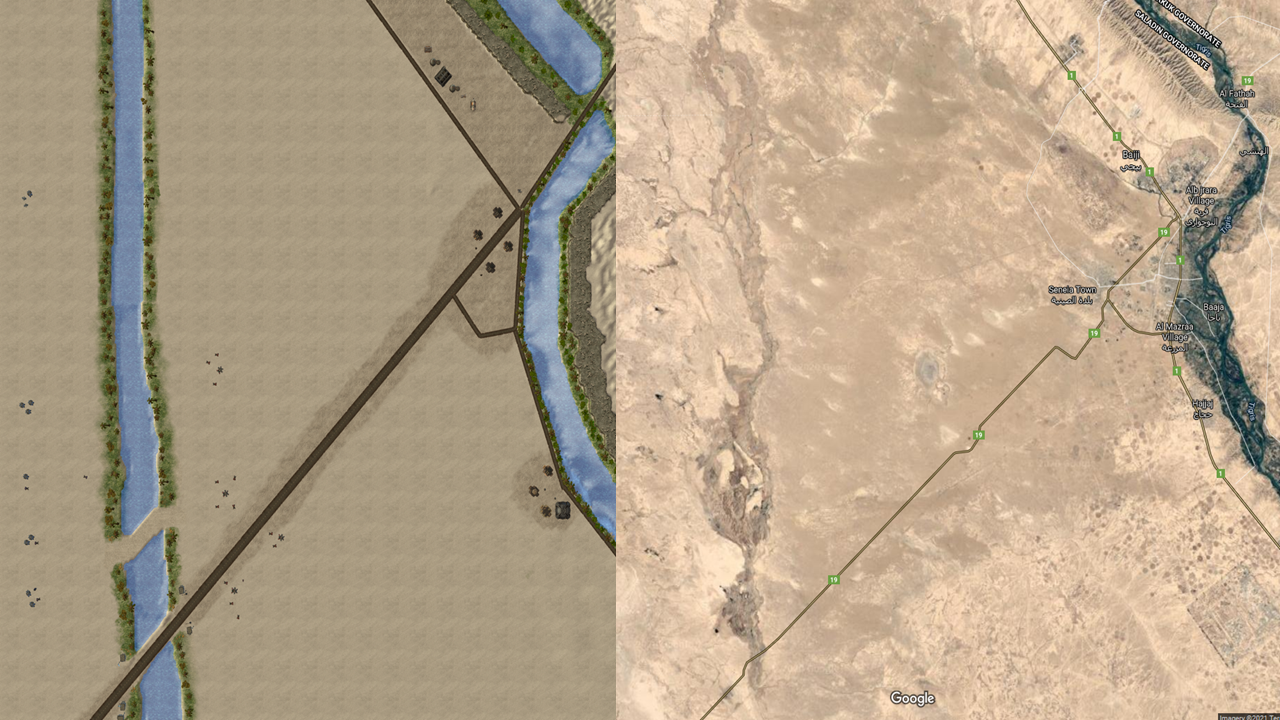}}
    \caption{TigerClaw: \textbf{(Right)} The geographical map of the scenario and \textbf{(Left)} the correspondingly designed map in StarCraft}
    \label{fig:TigerClaw-Comparison}
\end{figure}

\hfill

\vspace{0.5em}
\noindent
\textbf{TigerClaw.} The TigerClaw melee map (Figure~\ref{fig:TigerClaw-Comparison}) is a high-level recreation of the TigerClaw combat scenario (Figure~\ref{fig:TigerClaw-Scenario}) developed using the StarCraft II map editor. The scenario was developed by Army subject-matter experts (SMEs) at the Captain’s Career Course, Fort Moore, Georgia. 
The BlueForce is an Armored Task Force (TF) which consists of combat armor with M1A2 Abrams, mechanized infantry with Bradley Fighting Vehicles (BFV), mortar, armored recon cavalry with BFV, and combat aviation. The RedForce is a Battalion Tactical Group (BTG) with attached artillery battery and consists of mechanized infantry with BMP, mobile artillery, armored recon cavalry, combat aviation, anti-armor with anti-tank guided missiles (ATGM), and combat infantry. 

As seen in Figure~\ref{fig:TigerClaw-Scenario}, 
the BlueForce is a supporting effort with a mission to cross a dry riverbed (Wadi) and defeat the defending RedForce in preparation for a forward passage of lines by the main effort. The terrain is challenging in this scenario because there are only two viable wadi crossing points (Figure~\ref{fig:TigerClaw-Comparison}).

\begin{figure}[!ht]
    \centering
    \fbox{\includegraphics[width=0.8\linewidth]{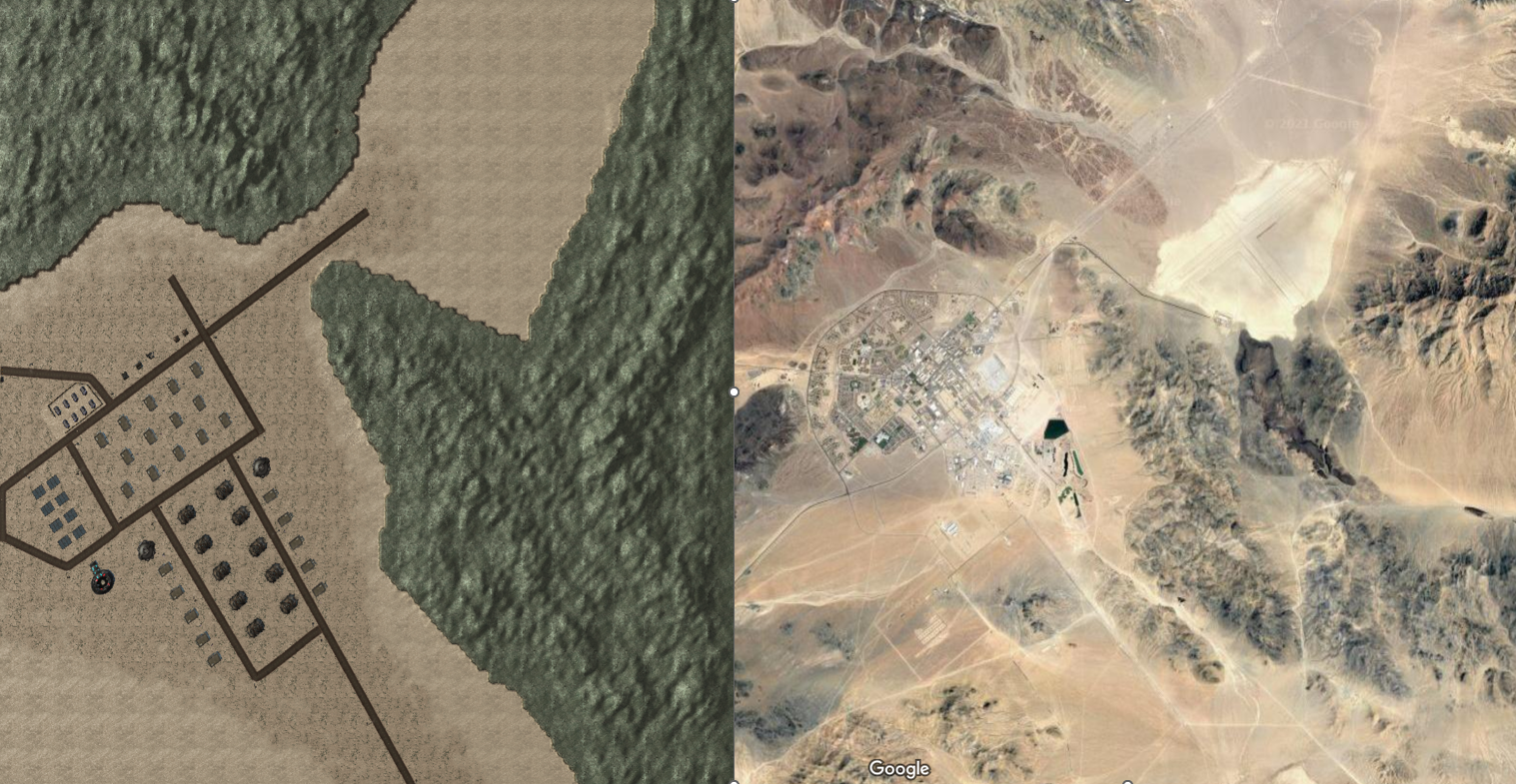}}
    \caption{NTC: \textbf{(Right)} The geographical map of the scenario and \textbf{(Left)} the correspondingly designed map in StarCraft}
    \label{fig:NTC-Comparison}
\end{figure}

% \hfill
\noindent
\textbf{National Training Center (NTC).} The NTC map 
is a representation of the Ft. Irwin and Bicycle Lake area as seen in Figure~\ref{fig:NTC-Comparison}.  The Blue and RedForce units are the same as TigerClaw, but there is also a capability to randomize the starting forces.  RedForce initially begins in the Bicycle Lake region, while BlueForce is set up in defense of Fort Irwin.  
Thus the RedForce will maneuver from the Bicycle Lake area and attempt to destroy the BlueForce in the Ft. Irwin area. The goal of this map was to investigate the impact of new terrain and to determine if the BlueForce would adopt a defensive strategy.

In both SC2 maps, each side has been represented down to the platoon echelon. The speed, attack range, and damage attributes of SC2 units have been scaled to estimate the capabilities of the relevant platforms in order to facilitate reinforcement learning.

\subsubsection{Reinforcement Learning Algorithms} 

\hfill

\noindent
\textbf{Asynchronous Advantage Actor-Critic~(A3C)~\cite{mnih2016asynchronous}. } A3C is an asynchronous version of the Advantage Actor-Critic (A2C) algorithm for RL that uses multiple agents to learn a policy and an estimate of the value function in parallel. 
At each timestep, the agent takes an action based on the current policy and receives a reward while transitioning to a new state. It then calculates the advantage function and updates its local copies of the actor-critic networks. This process takes place independently and asynchronously for each agent. A central parameter server that stores the global parameters of the networks is updated periodically by each agent and is used to initialize the local parameters of the agents' local networks.
The asynchronous nature of the updates leads to more efficient exploration of the environment and reduces the correlation between the updates leading to more stable and efficient learning.
Typically, the policy and value functions are parameterized by a shared neural network with a softmax output for the policy and a linear output for the value function.

\noindent
\textbf{Proximal Policy Optimization~(PPO)~\cite{schulman2017proximal}.} PPO is a policy gradient method that aims to improve the performance and stability of trust-region\cite{schulman2015trust} methods by introducing a \textit{clipped surrogate} objective function. This objective function effectively restricts policy change to a small range thus reducing the variability in training of the actor.
During training, the PPO algorithm iteratively updates the actor as well as the action value function and state value function using the temporal-difference~(TD) method. Further, this objective function enables PPO to guarantee monotonic improvements in the objective. 
This allows for faster convergence without strict constraints leading to more accurate and stable performance of the agent.

\section{RL Environment}
In this section we provide technical details of the RL environment used to train the C2 agent.
\subsection{State and Action Space} \label{sec:state_action}
The state space observed from the StarCraft II C2 environment consists of both a \emph{screen} or visual representation and a \emph{nonspatial} representation. The screen representation is a size $256\times 256$ image of the the minimap that depicts the current environmental state. 
The value of each pixel provides the agent with an understanding of the BlueForce and RedForce units' positions along with terrain information. The nonspatial representation consists of a vector of nonspatial features of size $287$ that encodes all of the game and unit information, such as unit type, health, position, game scores etc.     

The action space within the StarCraft II environment is large since it is a combination of the following three components: the number of units available to the commander, the number of possible actions that each unit can execute, and the $(x,y)$ pixel location within the size $256\times 256$ minimap where the action will be executed. To reduce the action space, the StarCraft II C2 environment first restricts the number of units by defining \emph{control groups}, which lumps common units together to reduce the overall number of units needed to be controlled. In our custom scenarios, the control groups are defined as -- \ballnumber{1} BlueForce: ``$\texttt{AVIATION}$", ``$\texttt{MECH\_INF}$", ``$\texttt{MORTAR}$", ``$\texttt{SCOUT}$" and  ``$\texttt{TANK}$" \ballnumber{2} RedForce: ``$\texttt{ANTI\_ARMOR}$", ``$\texttt{ARTILLERY}$", ``$\texttt{AVIATION}$", ``$\texttt{INFANTRY}$" and ``$\texttt{MECH\_INF}$"

Within each control group, we restrict the possible actions to be either ``$\texttt{NO\_OP}$" or ``$\texttt{ATTACK(x,y)}$", where the function of the former is to essentially do nothing, while the latter moves the control group (\ie all associated units) to a desired $(x,y)$ location and attacks any enemy within its firing range along the way. 
Additionally, the number of $(x,y)$ pixel locations is reduced by segmenting the minimap into nine disjoint quadrants with locations defined as ``$\textrm{LEFT}$", ``$\textrm{CENTER}$", and ``$\textrm{RIGHT}$" for the $x$-axis, and ``$\textrm{TOP}$", ``$\textrm{CENTER}$", and ``$\textrm{BOTTOM}$" for the $y$-axis, where the exact pixel location is the center of the quadrant.

\subsubsection{Reward Structure}
\label{sec:reward}

\hfill 

\noindent
\textbf{TigerClaw. }The reward function for the TigerClaw map consists of -
\begin{itemize}[wide, labelindent=10pt, itemsep=0pt]
    \item \textit{Terrain Reward:}  $+10$ points for each BlueForce unit crossing the Wadi (a dry river bed) and $-10$ points for retreating back.
    \item \textit{Attrition Reward:} $+10$ points for destroying a RedForce unit and $-10$ points if a BlueForce unit is destroyed.
\end{itemize}
The terrain reward is meant to reward an offensive strategy for the BlueForce. It is meant to reinforce crossing of the Wadi to initiate conflict with the RedForce in order to take over desired locations. To drive this scenario, the RedForce has been scripted to defend the desired locations.
\vspace{0.5em}

\noindent
\textbf{NTC. }The NTC map shares the same attrition reward function as TigerClaw, but the terrain rewards are not included. That is,
\begin{itemize}[wide, labelindent=10pt, itemsep=0pt]
    \item \textit{Attrition Reward:} $+10$ points for destroying a RedForce unit and $-10$ points if a BlueForce unit is destroyed.
\end{itemize}
The lack of terrain reward is meant to encourage the BlueForce to defend and focus on maximizing RedForce losses.  To drive this scenario the RedForce has been scripted to seek and destroy the BlueForce.

\subsection{RL Agent Description}
\subsubsection{The Policy Network}
\begin{figure}[h!]
    \centering
    \includegraphics[width=0.8\linewidth]{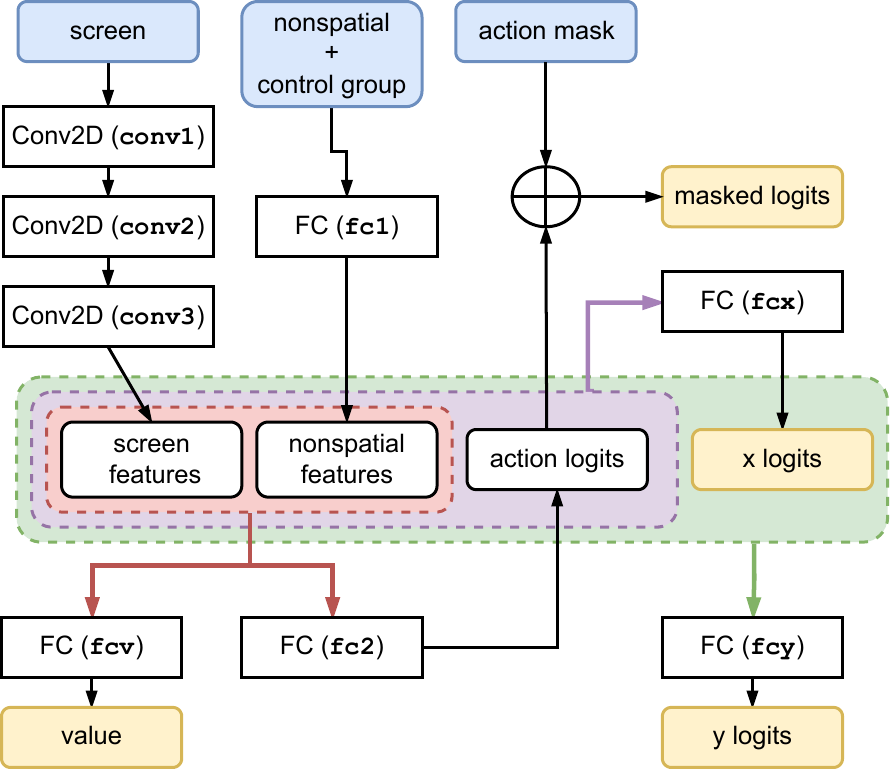}
    \caption{\textbf{C2 Policy Network: } Compuational graph of the policy network of our C2 agent. The inputs and outputs are shown in blue and yellow respectively. Shaded rectangles represent the \textit{concatenate} operation. Conv2D and FC layers are ReLU activated.}
    \label{fig:policy-network}
\end{figure}
\hfill
\hfill

\noindent
The C2 agent in this paper uses the same policy network as our previous work~\cite{waytowich2022learning}. The network takes as inputs, three kinds of observations and the control group
\begin{itemize}[leftmargin=*]
    \item \textit{screen} - The screen representation discussed in Section~\ref{sec:state_action} consisting of a vector of an image of size $256\times 256$. 
    
    \item \textit{nonspatial} - The nonspatial representation discussed in Section~\ref{sec:state_action} consisting of a vector of size $287$. 
    
    \item \textit{action mask} - The action mask is used to restrict the action space to the allowed actions as described in Section~\ref{sec:state_action}. 
    
    \item \emph{control group} - A one-hot encoding of the selected control group (specified in Section~\ref{sec:state_action}) that will be the focus of action prediction by the policy network. The encoding is then concatenated to the nonspatial input vector.
    At each time step, the control group is sequentially selected to identify their next actions.
    
\end{itemize}

The output of the network is an approximation of the value function and an $8$ element vector arranged as \textit{(action-logits, x-logits, y-logits)}. 
The \textit{action-logits} is a $2$ element vector that determines the action.
The \textit{x,y-logits} are $3$ dimensional entries 
each with the logit values corresponding to the positions $(\textrm{LEFT},\textrm{CENTER},\textrm{RIGHT})$ and $(\textrm{TOP},\textrm{CENTER},\textrm{BOTTOM})$ on the map, respectively. 
This output vector is used to create a probability distribution over the action space from which the next action is sampled by the agent.

\section{Adversarial Attacks on RL agents}
\label{sec:technical}

Prior works on the robustness of RL training have focused on evaluating the algorithms from a perspective of sensitivity to environment dynamics \cite{huang2017adversarial,sun2020stealthy} or the ability to train adaptive adversarial policies against them \cite{wu2021adversarial,gleave2019adversarial}. In this work, we focus on the former approach. 
A lot of previous research has shown that neural network predictions are highly sensitive to perturbations in their input space\cite{7958570,goodfellow2014explaining,madry2018towards}. 
As DRL approaches typically rely on parameterizing policies with neural networks, they suffer from the same vulnerabilities.

\subsection{Adversarial Attacks on Image Classifiers} 
\label{sec:adv-img-classifier}

An adversarial perturbation is a small perturbation that is added to a benign input to fool a trained network into predicting an incorrect output. Typically the perturbation is constrained to be small enough to avoid detection. For example, in the case of image classifiers, the ideal adversarial perturbations in pixel space would be imperceptible to a human observer but would cause the neural network classifier to predict an incorrect class with high confidence. Formally, for a trained classifier $f(w;.)$ and an input image $x$, computing an adversarial sample $x' = x + \sigma$ involves computing $\sigma$ such that $f(w,x') \neq f(w, x)$ while minimizing $d(x', x)$, where $d$ is a distance metric such as Euclidean distance.

\subsubsection{Fast Gradient Sign Method (FGSM)~\cite{goodfellow2014explaining}. } 
FGSM provides an efficient method for generating adversarial samples given whitebox access to the model. Given a trained classifier model $f(w;.)$, a first order attacker can generate adversarial perturbations for the benign sample $x$ by first computing the gradient of the classifier loss ($L$) with respect to the input. The weighted perturbation when added to $x$, creates the adversarial sample $x'$. 
$$x' = x + \varepsilon \cdot\text{sgn}(\nabla_x L(f(w;x), y_{t}))$$ 
where $y_{t}$ is the ground truth label of the sample $x$ and $\varepsilon$ is the perturbation budget that controls the amount of distortion to the original sample. In the case of FGSM, it bounds the $l_{\infty}$ norm of the perturbation added to the original image. Intuitively, the attack moves the sample $x$ in the direction of $\nabla_x L(f(w;x), y_{t})$  which maximizes the classifier loss $L$. In this work, we prefer FGSM to more powerful attacks such as PGD\cite{madry2018towards} and C\&W\cite{7958570} because of its lower computational cost, which leads to a more efficient attack.

\begin{figure*}[!ht]
    \centering
    % \hfill
    \begin{subfigure}{0.24\linewidth}
        \includegraphics[width=\linewidth]{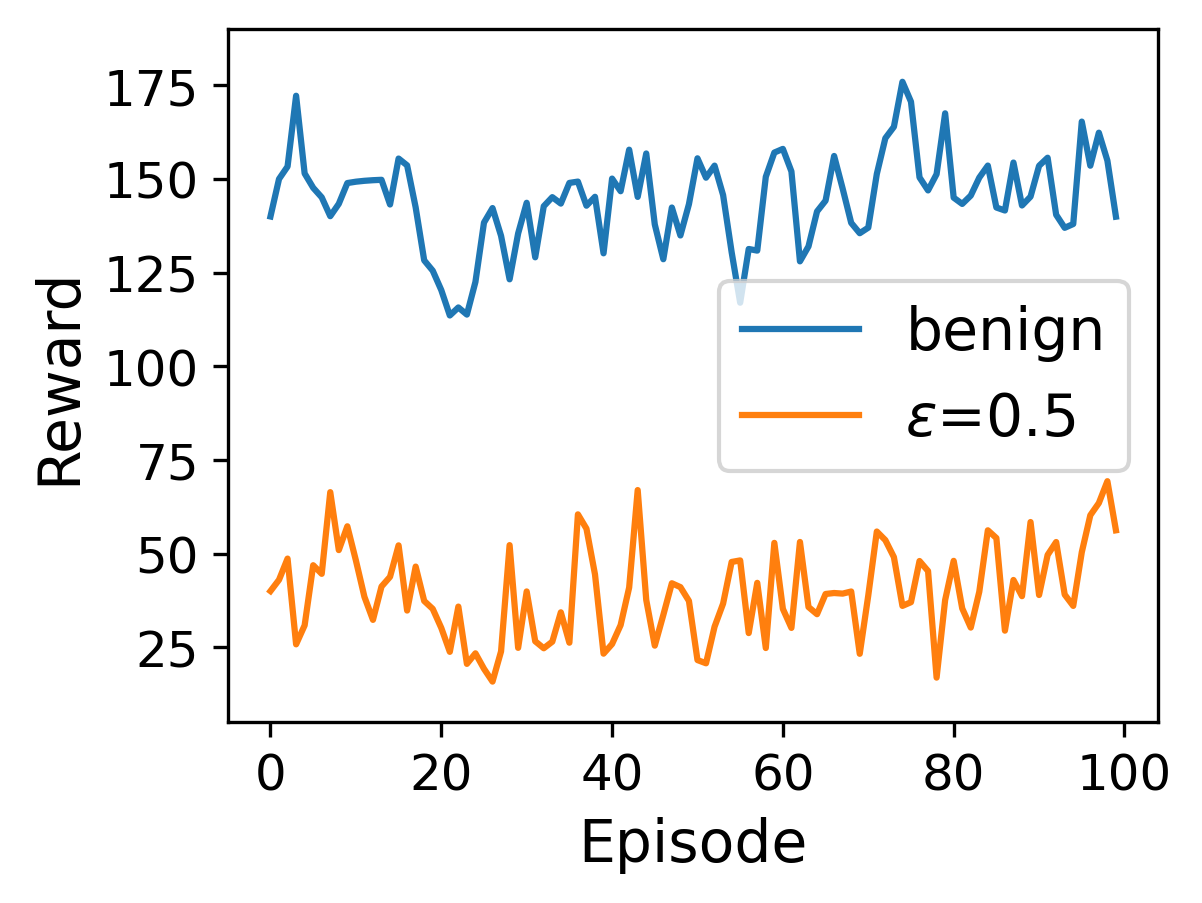}
        \caption{PPO/TigerClaw: Episode Rewards over 100 episodes}
        \label{fig:inf-PPO-1}
    \end{subfigure}
    \begin{subfigure}{0.24\linewidth}
        \includegraphics[width=\linewidth]{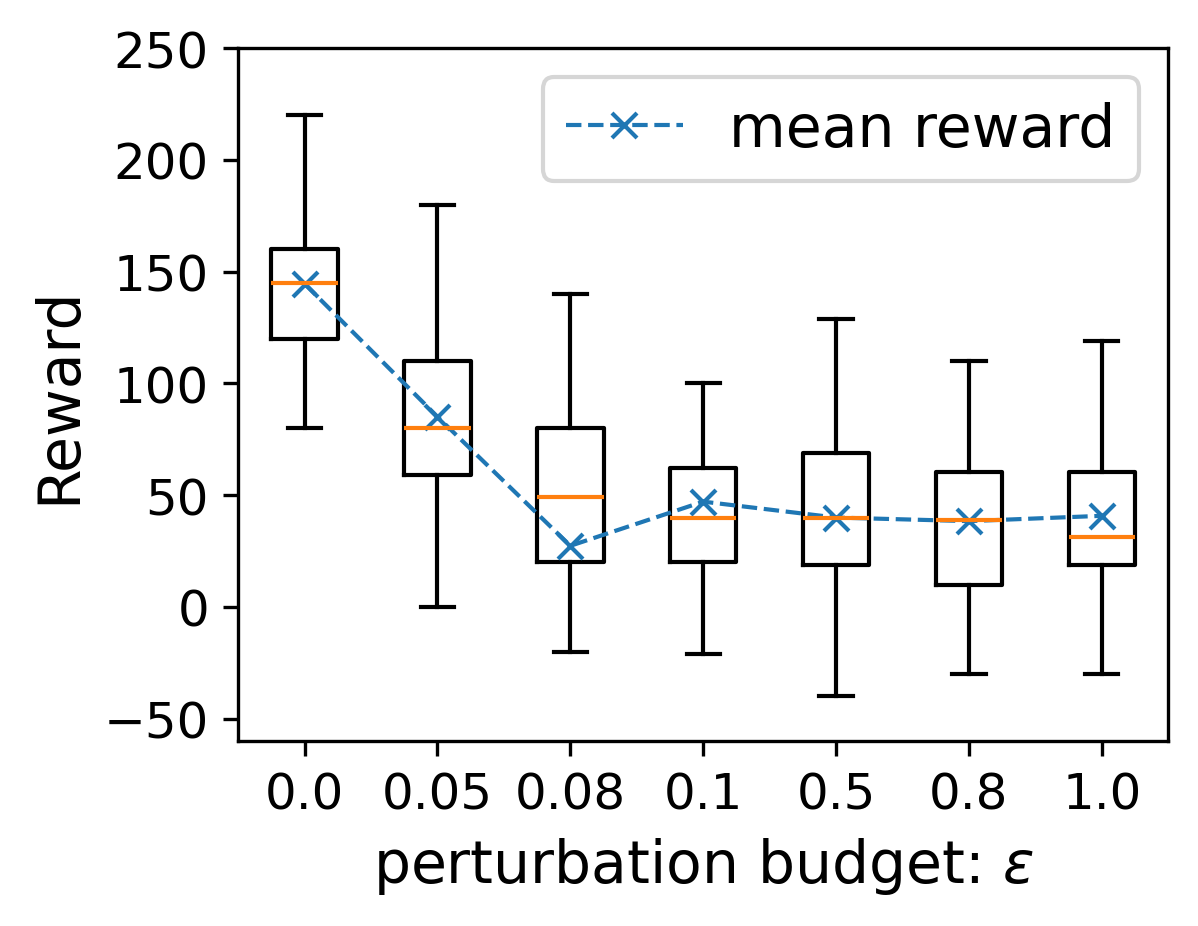}
        \caption{PPO/TigerClaw: Reward trend w.r.t perturbation budget $(\varepsilon)$.}
        \label{fig:inf-PPO-2}
    \end{subfigure}
    \begin{subfigure}{0.24\linewidth}
        \includegraphics[width=\linewidth]{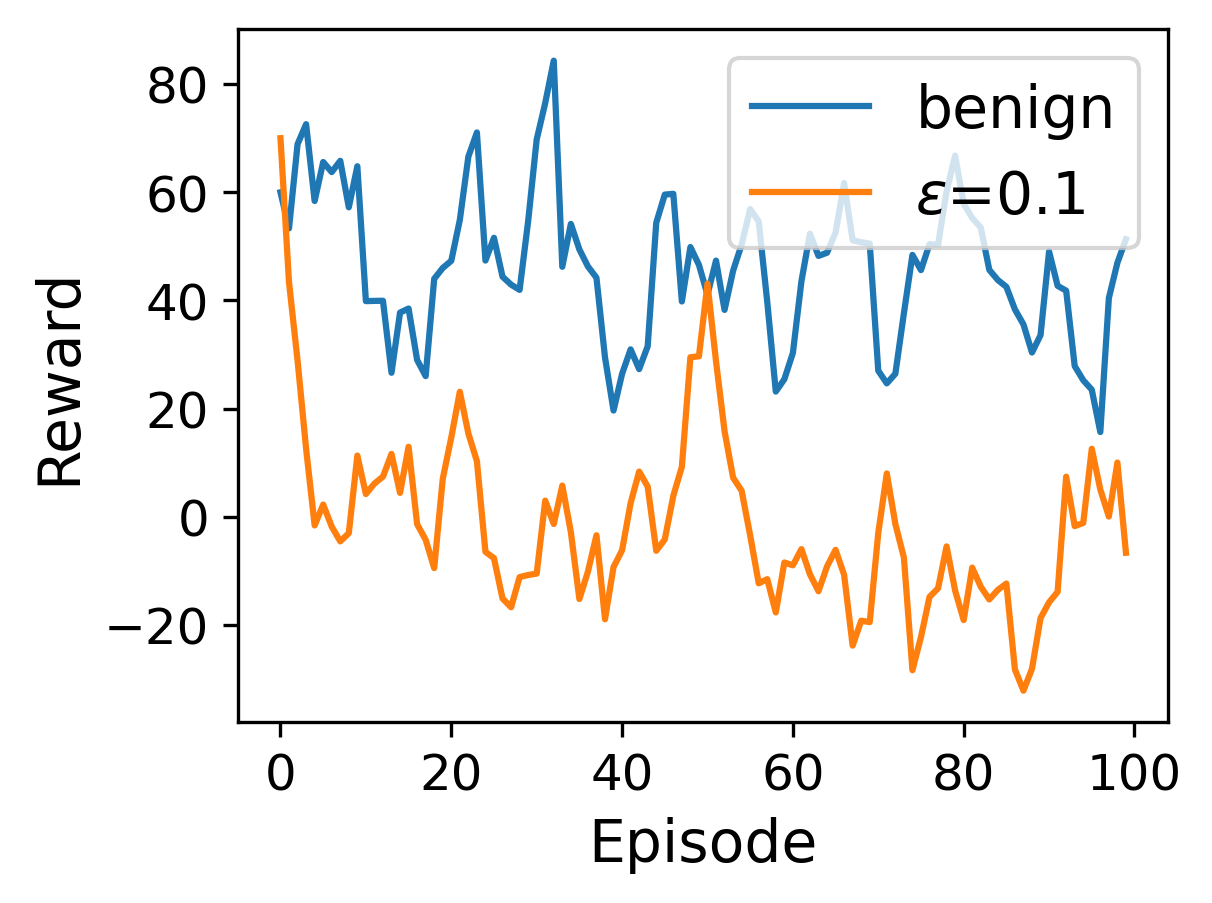}
        \caption{PPO/NTC: Episode Rewards over 100 episodes}
        \label{fig:inf-PPO-3}
    \end{subfigure}
    \begin{subfigure}{0.24\linewidth}
        \includegraphics[width=\linewidth]{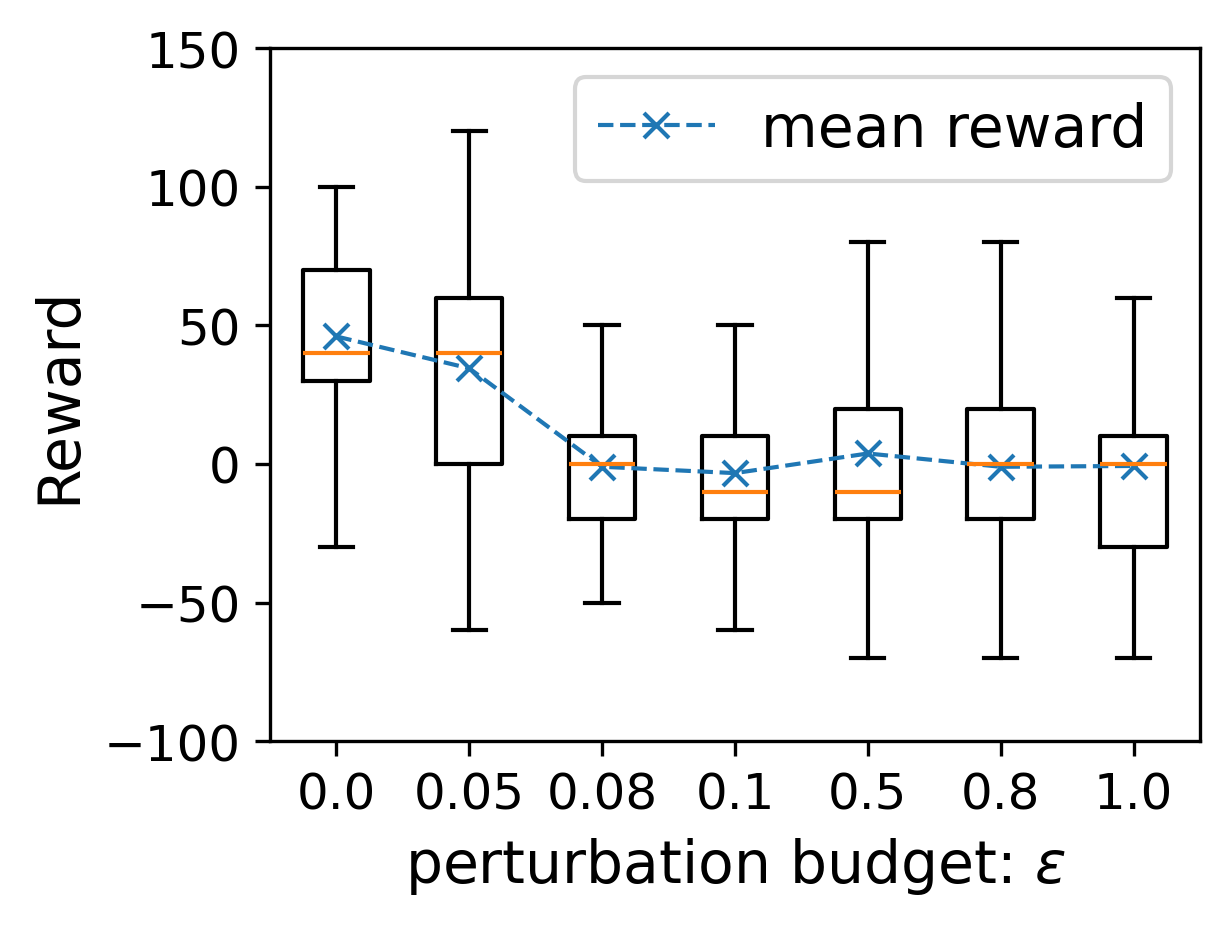}
        \caption{PPO/NTC: Reward trend w.r.t perturbation budget $(\varepsilon)$.}
        \label{fig:inf-PPO-4}
    \end{subfigure}
    \caption{Inference time attack on an agent in the TigerClaw~(a),(b) and NTC~(c),(d) scenario trained using PPO.}
    \label{fig:inf-PPO}
\end{figure*}

\begin{figure*}[!ht]
    \centering
    % \hfill
    \begin{subfigure}{0.24\linewidth}
        \includegraphics[width=\linewidth]{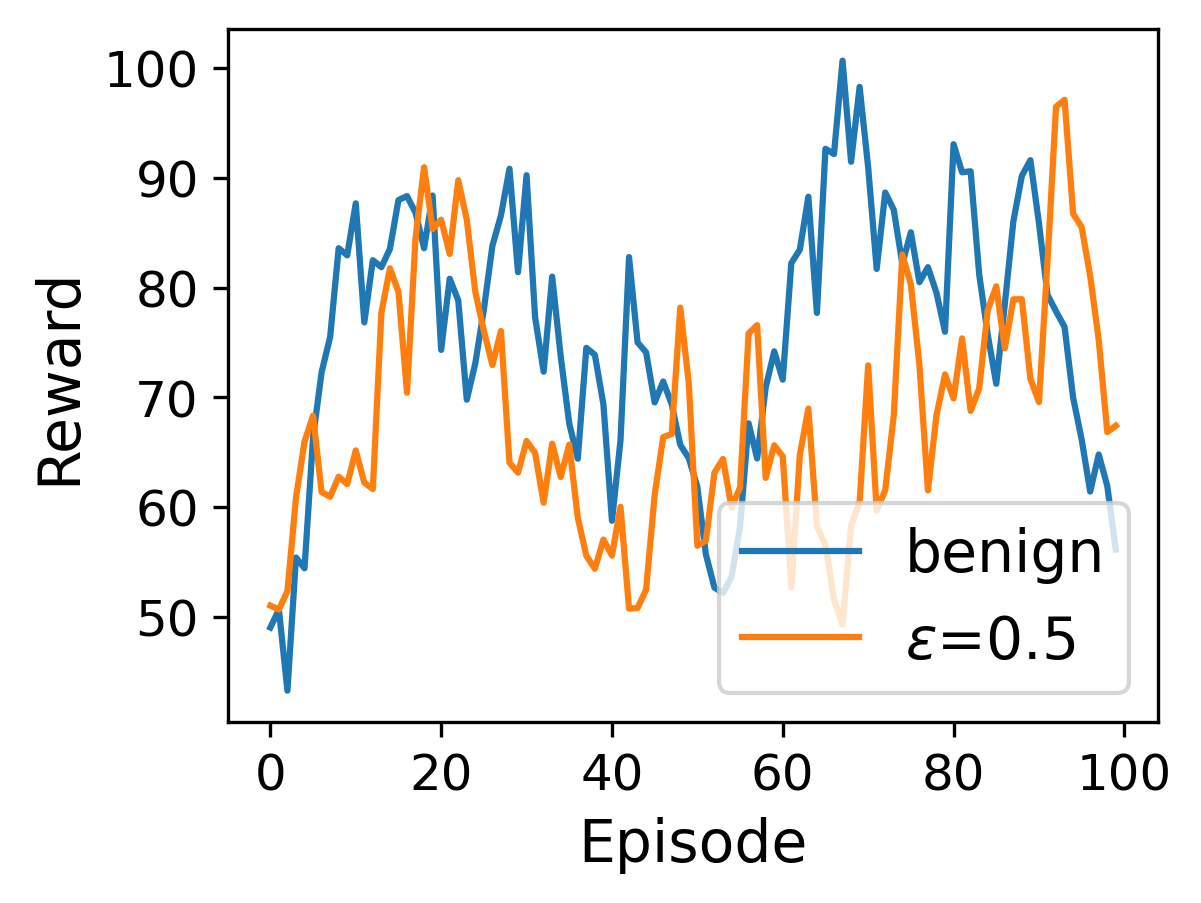}
        \caption{A3C/TigerClaw: Episode Rewards over 100 episodes}
        \label{fig:inf-A3C-1}
    \end{subfigure}
    \begin{subfigure}{0.24\linewidth}
        \includegraphics[width=\linewidth]{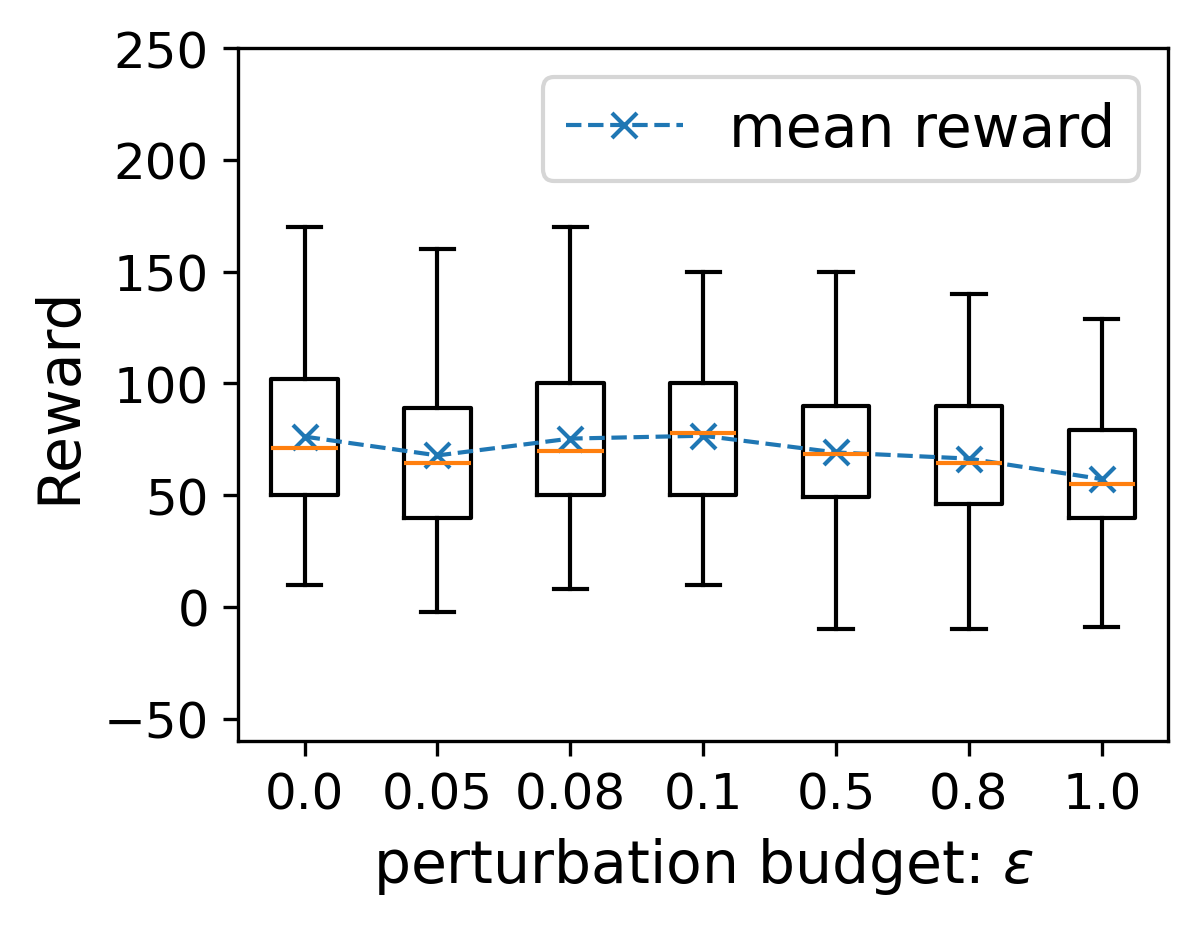}
        \caption{A3C/TigerClaw: Reward trend w.r.t perturbation budget $(\varepsilon)$.}
        \label{fig:inf-A3C-2}
    \end{subfigure}
    \begin{subfigure}{0.24\linewidth}
        \includegraphics[width=\linewidth]{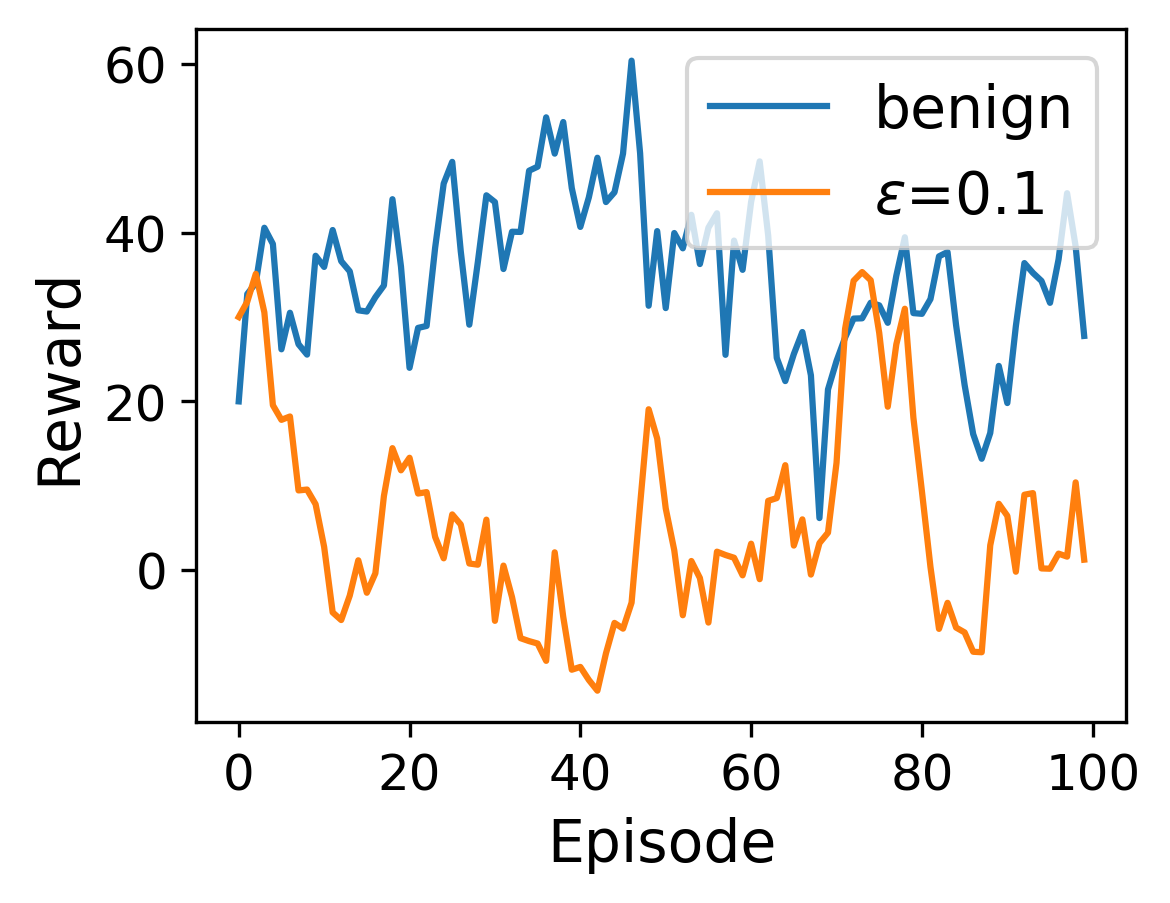}
        \caption{A3C/NTC: Episode Rewards over 100 episodes}
        \label{fig:inf-A3C-3}
    \end{subfigure}
    \begin{subfigure}{0.24\linewidth}
        \includegraphics[width=\linewidth]{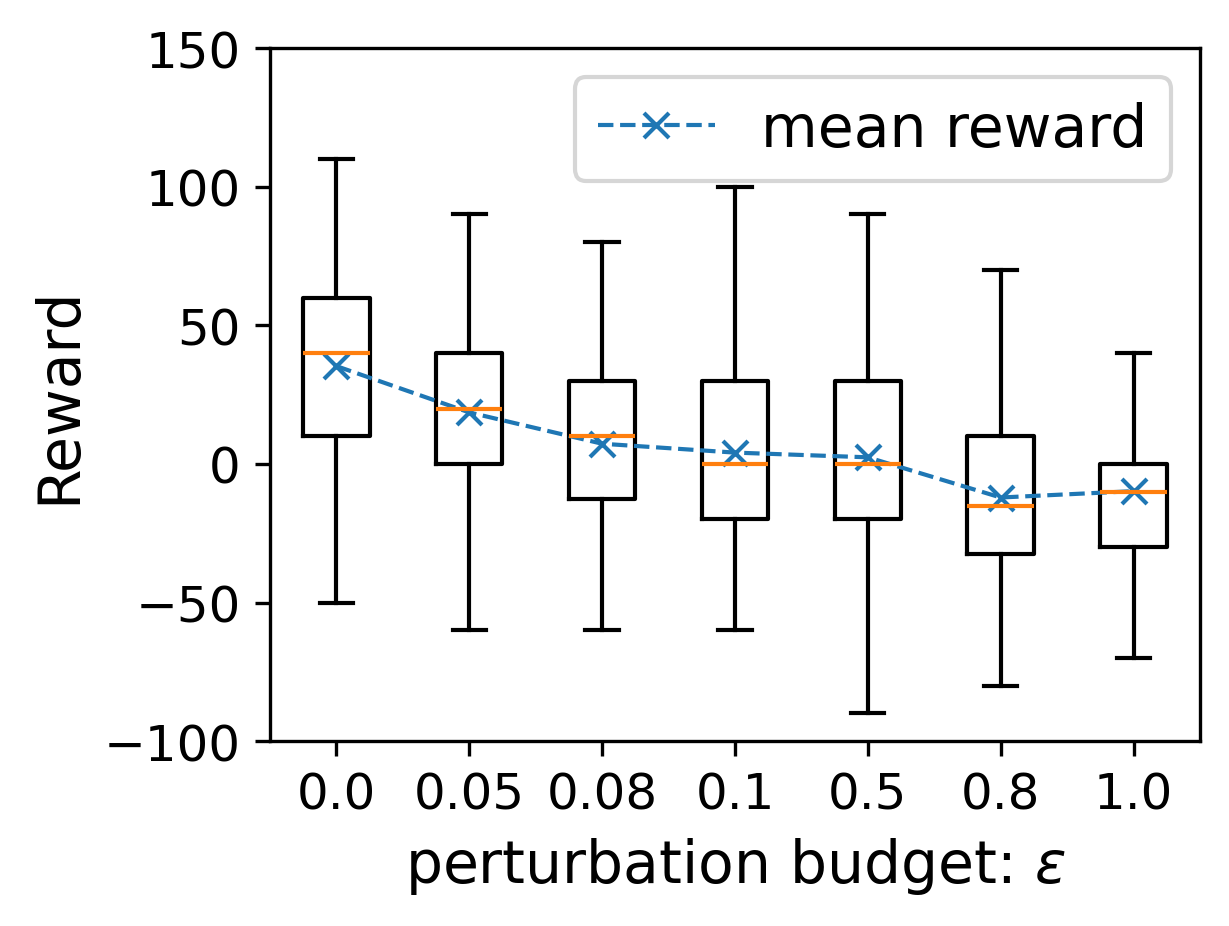}
        \caption{A3C/NTC: Reward trend w.r.t perturbation budget $(\varepsilon)$.}
        \label{fig:inf-A3C-4}
    \end{subfigure}
    \caption{Inference time attack on an agent in the TigerClaw~(a),(b) and NTC~(c),(d) scenario trained using A3C.}
    \label{fig:inf-A3C}
\end{figure*}

\subsection{Inference Time Attacks on Policies} 
\label{sec:inf-time-attack}

Inference time attacks are used against pre-trained agents which are deployed in the wild. The goal of the attack is to induce the trained policy network into predicting a sub-optimal action distribution by surreptitiously perturbing the input observations. This inevitably leads to the agent losing its expected reward. 

In the wild, such an attack could be realized as a cyber-attack where sensors collecting data from the battlefield could be compromised and transmit corrupted data. 
A high level overview of the attack is presented in Figure~\ref{fig:intro-fig}.

\subsubsection{Threat Model. }
First we assume a C2 agent that has been trained in a benign environment and has learned an optimum policy. This agent is subsequently deployed in an unsafe setting to direct BlueForce troops in a battle against the RedForce.
Next, we consider an attacker that has the ability to intercept and modify observations coming from the environment before they are received by the C2 agent to select the next actions. 
Through such modifications, the attacker hopes to influence the agent to sample an incorrect action that leads to a low reward for the BlueForce. To generate such input perturbations efficiently, we also assume that the attacker has white-box access to the policy network used by the C2 agent.

\subsubsection{Attack Methodology. }
We use FGSM as a basis for generating perturbations at inference time and use the modification presented by Huang \etal \cite{huang2017adversarial} in order to target policies instead of classifiers.
Unlike supervised learning, %since for RL 
we do not know the ``ground truth" action at any given timestep in RL and we assume that the action predicted by the policy network with the highest likelihood is optimal.

It should be noted that since the output of our policy network has three different components (action logits, x-logits, y-logits), 
ideally the ground truth vector should be computed for each component separately.
With this in mind we construct the ground truth vector with a \textit{degenerate distribution}. That is, for an input observation $x$, for each component of the output, we take the element with the highest value in the vector $y = [y^{(1)}, y^{(2)}, y^{(3)}]=f(w;x)$ and assign weight $1$ to it's logit value and $0$'s everywhere else. In other words,
\begin{align*}
y'^{(j)}_i=\left\{
  \begin{array}{@{}ll@{}}
    1, & \text{if}\ y^{(j)}_i = \max(y^{(j)}) \\
    0, & \text{otherwise}
  \end{array}\right.
\end{align*}
where $y_i$ is the $i^{th}$ element of vector $y$. 

However, we find that treating the components as a single unit and calculating the degenerate distribution on the entire output is more efficient while making for an effective attack. After making this relaxation, the malicious perturbation is then calculated for $x$ using the gradient $\nabla_x L(f(w; x) ,y')$, where $L$ is the cross-entropy (CE) loss function.
We use this technique to craft perturbations for two of the three input components namely -- the \textit{screen} and \textit{nonspatial} components. We do not perturb the action mask as it is simply used to mask out invalid actions predicted by the network. As a result, perturbing this component does not conform to any realistic setting.

\section{Evaluations}
\label{sec:eval}
\subsection{Experimental Setup}
Our experiments were conducted on a cluster node with two AMD Epyc 7763 ``Milan" CPUs 128 cores with 256GBs of main memory. We used the SC2 framework described in Section~\ref{sec:sc2env} to train our RL agents. Each training run was performed using 90 workers, each occupying a single core (90 parallel instances of the game) and consumed about 170GBs of memory. The PPO and A3C agents were trained for 5K/25K iterations respectively corresponding to around 40M/100M timesteps or 134K/376K episodes respectively. 
The rollouts were visualized using a custom \texttt{pygame} interface.

\subsection{Evaluation of Inference Time Attacks on Agent Reward}
For evaluating  effectiveness of inference time attacks, first we train our C2 agent in a benign environment until it learns a policy that consistently achieves a high reward on the given scenario. 
The trend of the attained reward when the agent is deployed in the presence of an attacker is then studied over the course of a 100 episodes or \textit{rollouts}.

For a comprehensive evaluation, we consider two scenarios or tasks -- \ballnumber{1} TigerClaw (Attacking BlueForce) and \ballnumber{2} NTC (Defending BlueForce). We also consider two different state-of-the-art training algorithms, A3C and PPO. In both scenarios the C2 agent controls the movements of the BlueForce while the RedForce follows a fixed policy. A detailed description of these scenarios was given in Section~\ref{sec:sc2env}.

To further analyze and understand the shift in agent strategy we use several quantitative metrics as well as observe multiple rollouts. Our insights are presented below.

\subsubsection{Vulnerability to Adversarial Perturbations. }

Figures~\ref{fig:inf-PPO} and~\ref{fig:inf-A3C} show the resulting reward trends for an agent trained using PPO and A3C, respectively. 
We present boxplot statistics aggregated over 100 episodes, for the rewards attained by the agent under attack with different perturbation levels in Figures~\ref{fig:inf-PPO-2} and~\ref{fig:inf-PPO-4} (PPO) and Figures~\ref{fig:inf-A3C-2} and~\ref{fig:inf-A3C-4} (A3C). 
$\varepsilon=0.0$ corresponds to the benign case when no perturbations are made.
In most cases, we observe a steep decrease in the median reward even for minute perturbations ($\varepsilon=0.05, 0.08$). Increasing $\varepsilon$ also shows diminishing returns for attack effectiveness. 

\begin{figure}[!ht]
    \centering
    \includegraphics[width=0.80\linewidth]{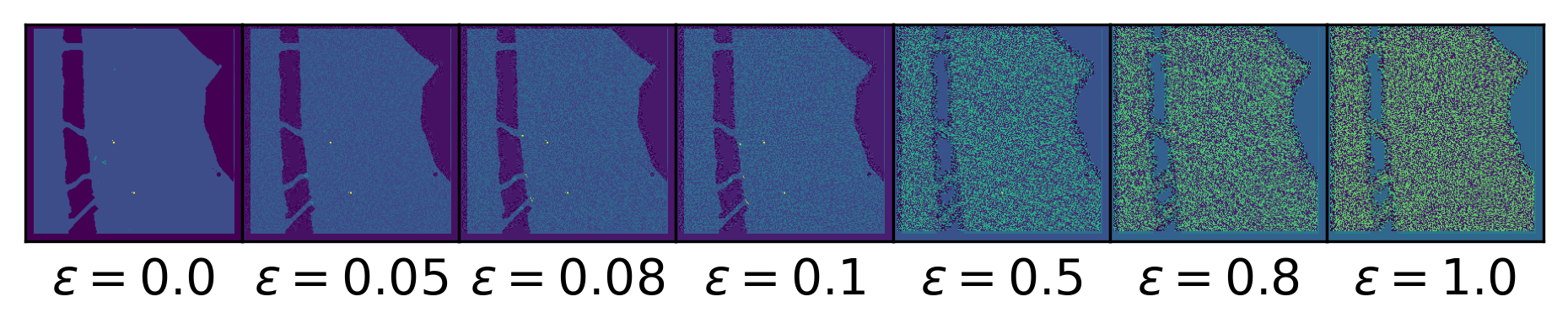}
    \caption{Visualizing perturbations of the \texttt{screen} component at different $\varepsilon$ levels in the TigerClaw scenario. Higher $\varepsilon$ levels result in a greater amount of noise. 
    }
    \label{fig:mmap-cmbd}
\end{figure}

To maintain the secrecy of the attack, the perturbation budget needs to be small enough to be imperceptible to a human auditor, especially for the screen component of the input. Figure~\ref{fig:mmap-cmbd} shows a visual comparison of perturbations of the screen component at different levels. From our evaluations we observe that $\varepsilon = 0.1$ is optimal in lowering the reward while keeping perturbations to a minimum.

\begin{figure}[!ht]
\captionsetup[subfigure]{justification=centering}
    \centering
        ${timestep}=10$\\
        \begin{subfigure}{0.44\linewidth}
        \includegraphics[width=\linewidth]{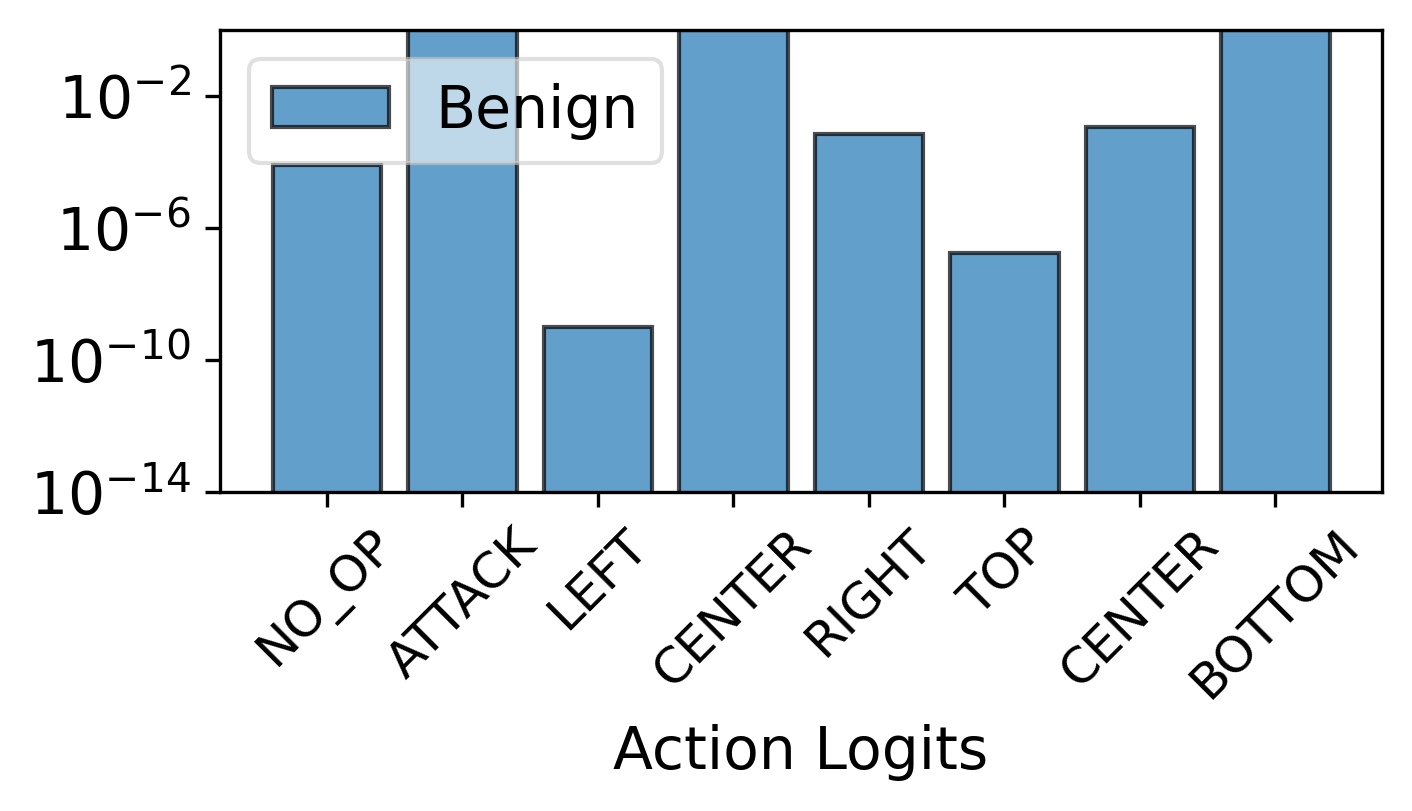}
    \caption{Predicted Action: \texttt{ATTACK(CENTER,BOTTOM)}}
        \label{fig:dist-comp-1}
    \end{subfigure}
    \begin{subfigure}{0.44\linewidth}
        \includegraphics[width=\linewidth]{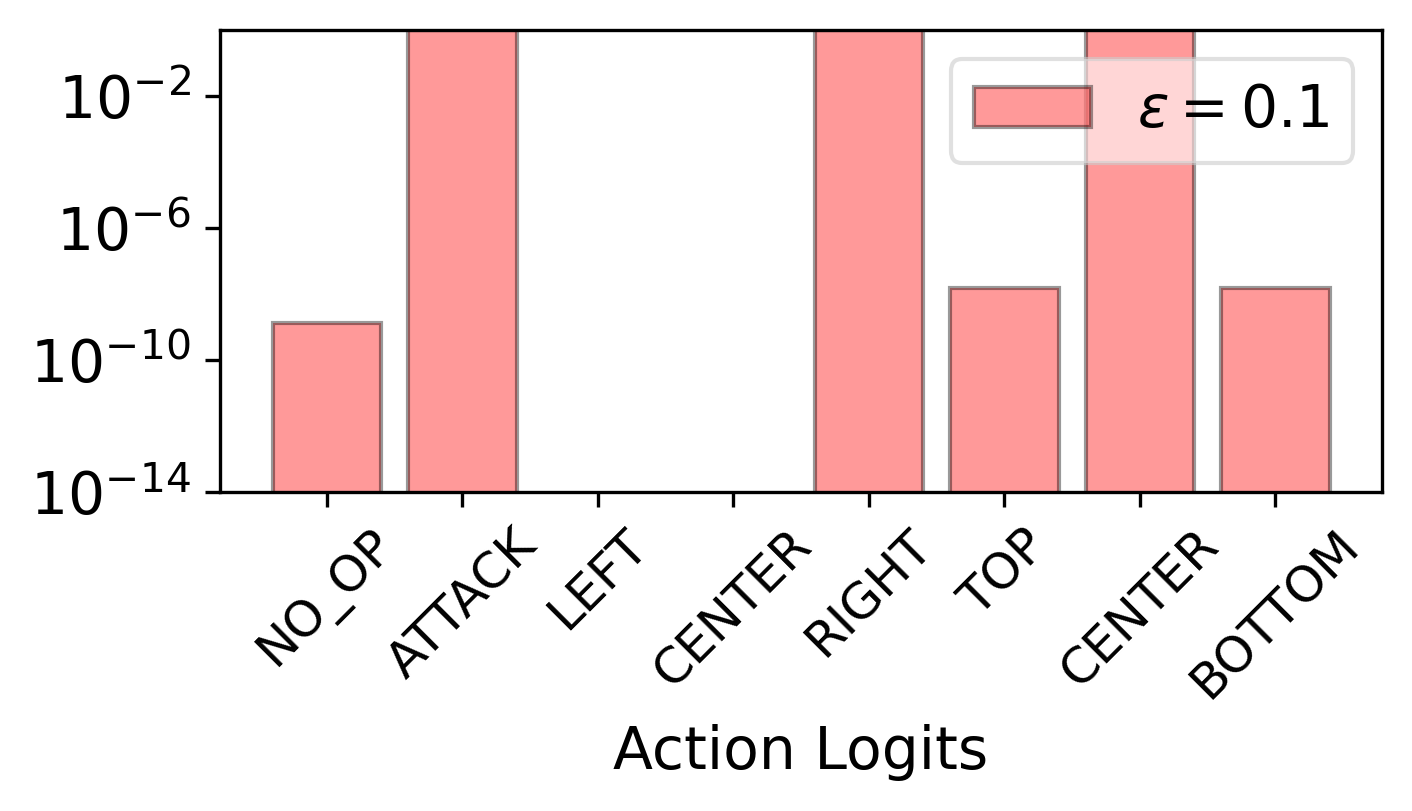}
    \caption{Predicted Action: \texttt{ATTACK(RIGHT,CENTER)}}
        \label{fig:dist-comp-2}
    \end{subfigure}
    \vspace{0.5em}\\$timestep=21$\\
    \begin{subfigure}{0.44\linewidth}
        \includegraphics[width=\linewidth]{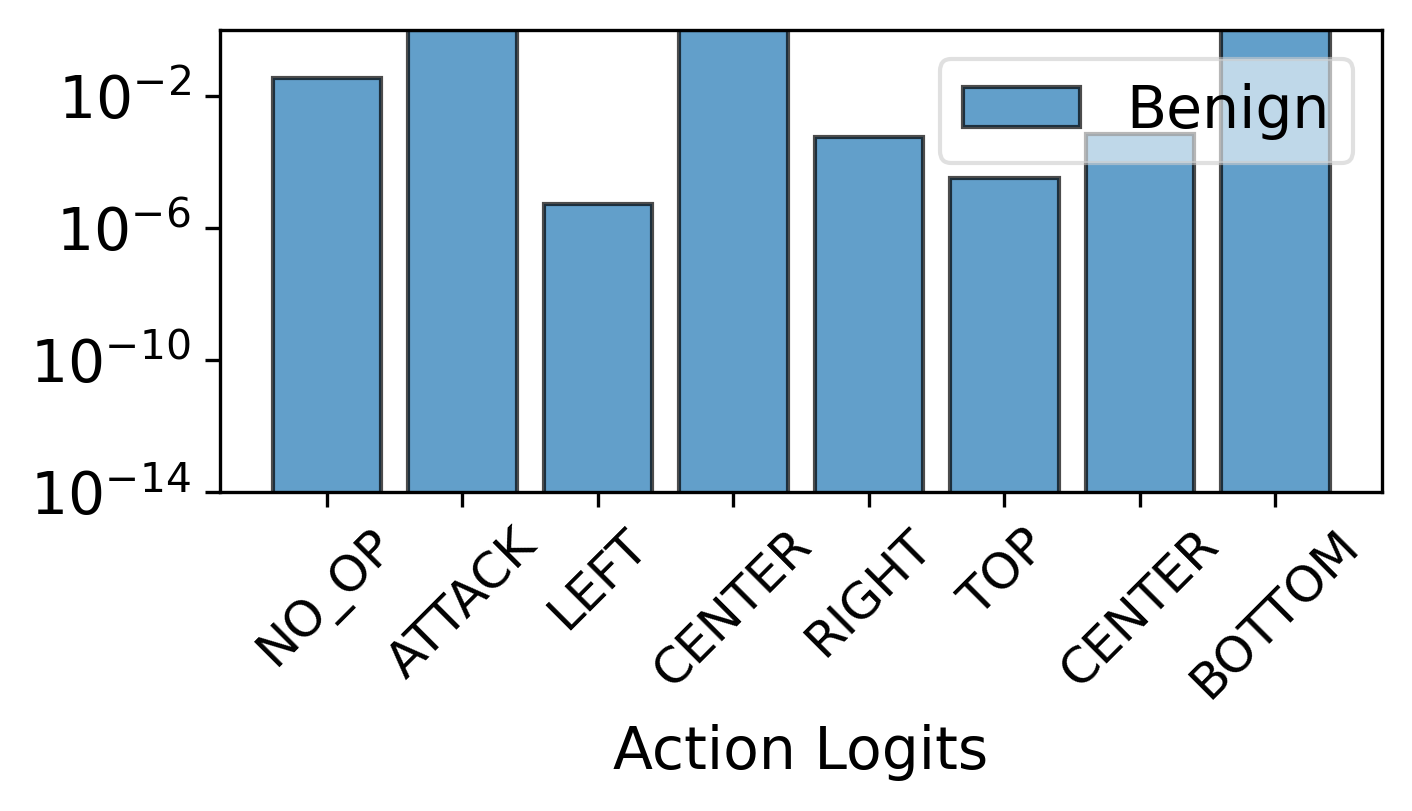}
        \caption{Predicted Action: \texttt{ATTACK(CENTER,BOTTOM)}}
        \label{fig:dist-comp-3}
    \end{subfigure}
    \begin{subfigure}{0.44\linewidth}
        \includegraphics[width=\linewidth]{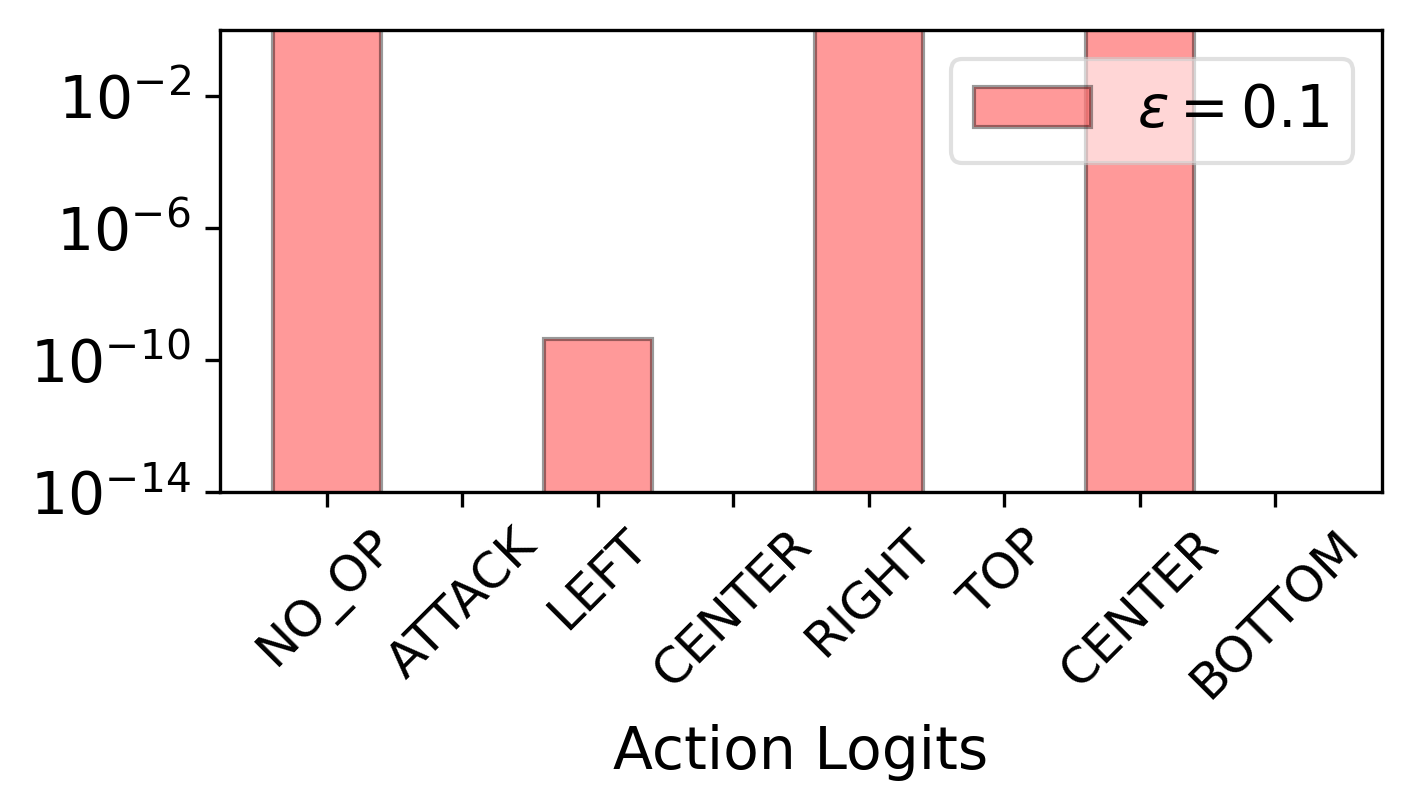}
        \caption{Predicted Action:\\ \texttt{NO\_OP}}
        \label{fig:dist-comp-4}
    \end{subfigure}
    \vspace{0.5em}\\$timestep=100$\\
    \begin{subfigure}{0.44\linewidth}
        \includegraphics[width=\linewidth]{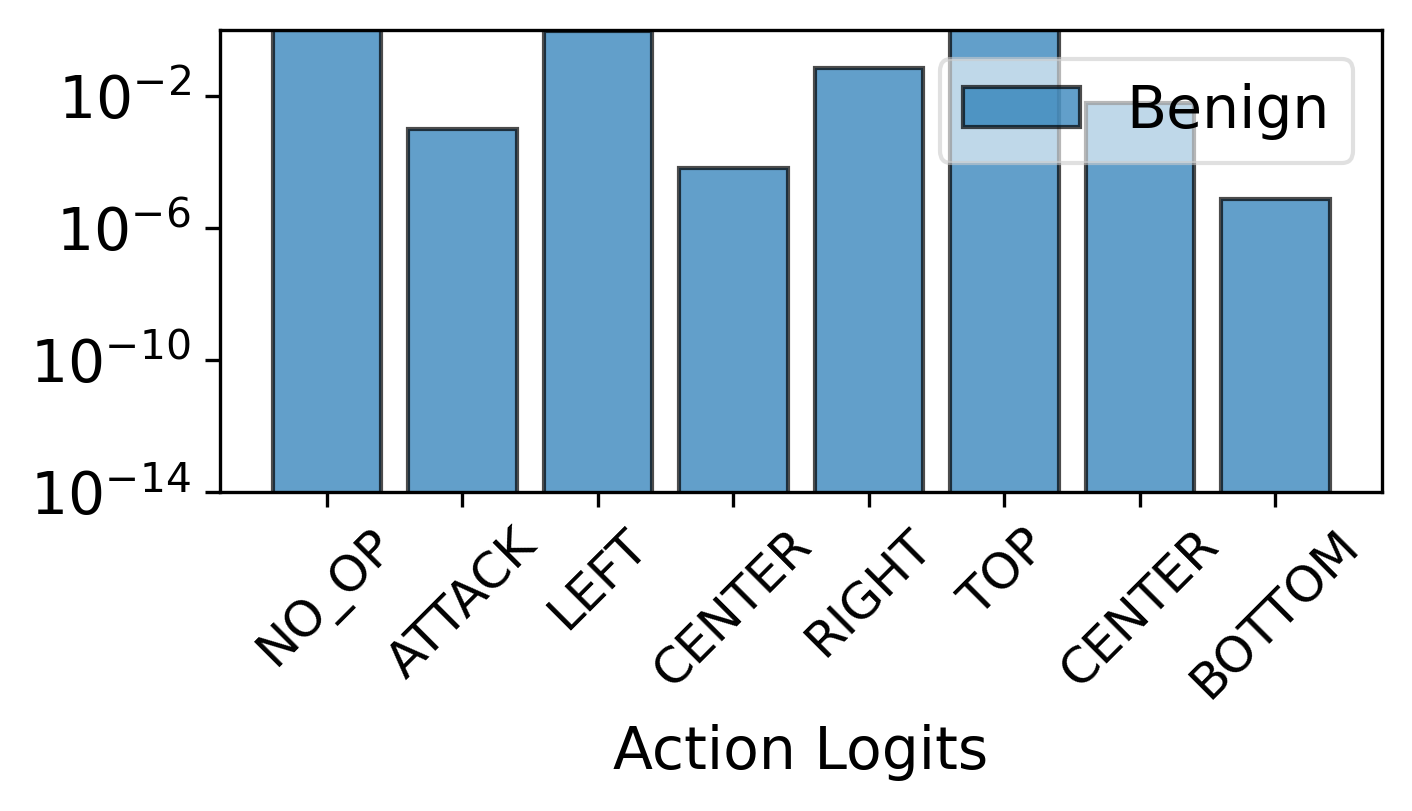}
        \caption{Predicted Action:\\ \texttt{NO\_OP}}
        \label{fig:dist-comp-5}
    \end{subfigure}
    \begin{subfigure}{0.44\linewidth}
        \includegraphics[width=\linewidth]{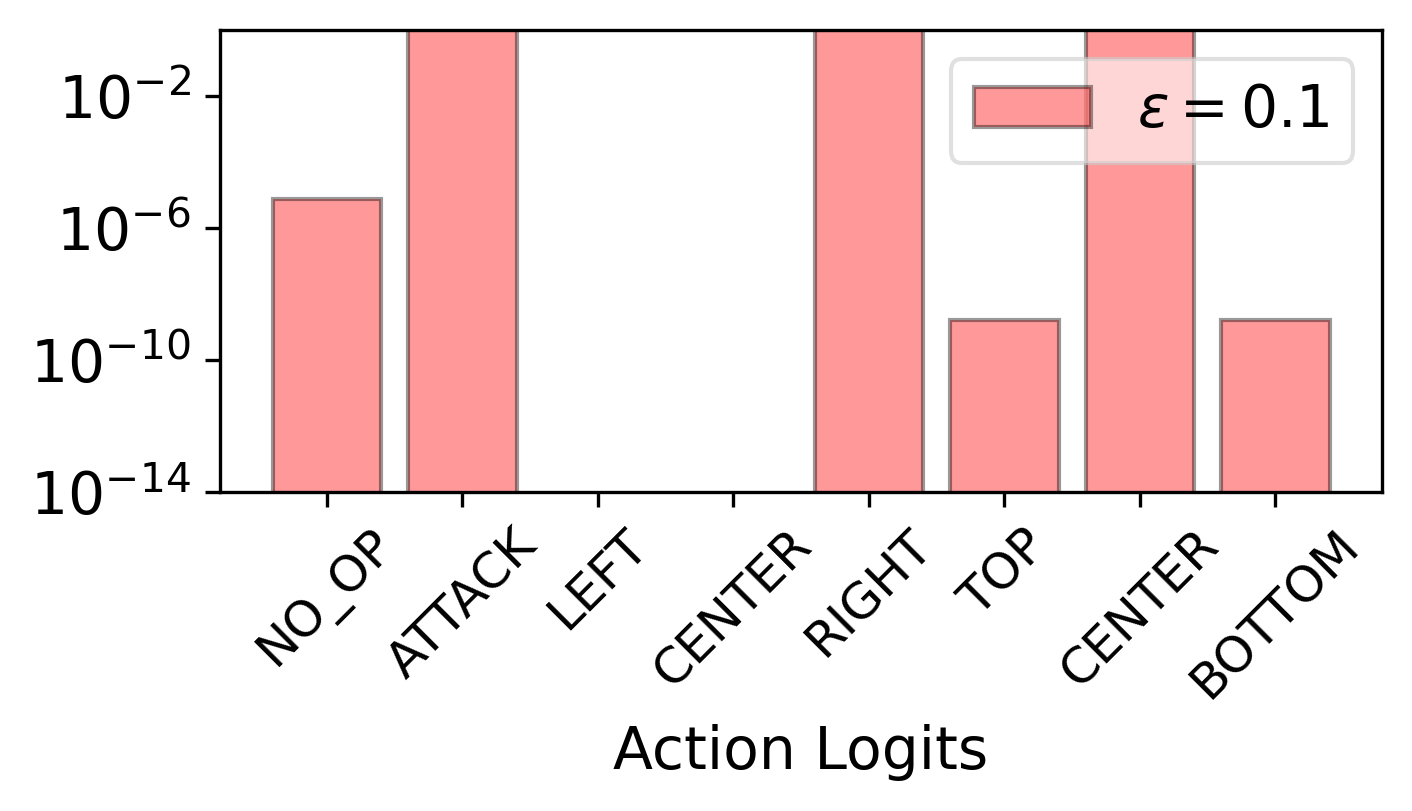}
    \caption{Predicted Action: \texttt{ATTACK(RIGHT,CENTER)}}
        \label{fig:dist-comp-6}
    \end{subfigure}
    \caption{Shift in the action distributions predicted by the policy network at different timesteps during a PPO/TigerClaw rollout. \textbf{(Left)} Action distribution on benign observations. \textbf{(Right)} Action distribution predicted by network after perturbing observation with $\varepsilon=0.1$. Each caption represents the action sampled with highest likelihood.}
    \label{fig:dist-comp}
\end{figure}

Figure~\ref{fig:dist-comp} provides explainability into the utility of the attack by showing how it changes the actions taken by the agent. We first plot the action distribution predicted by the policy network on benign observations collected at certain timesteps during a PPO/TigerClaw rollout.
These are shown in Figure~\ref{fig:dist-comp-1}, \ref{fig:dist-comp-3}, and~\ref{fig:dist-comp-5}, respectively. We then compare it to the distribution predicted by the same policy network after maliciously perturbing the observations with our attack for $\varepsilon = 0.1$ in Figures~\ref{fig:dist-comp-2}, \ref{fig:dist-comp-4}, and~\ref{fig:dist-comp-6}, respectively. The probabilities are plotted on a log scale. As can be observed the actions with the highest likelihood shift from  $(1,1,2) \rightarrow (1,2,1), (1,1,2) \rightarrow (0,2,1)~\&~ (0,0,0) \rightarrow (1,2,1) $ at timestep 10, 21, and 100. 

\subsubsection{Analyzing Agent Behavior under Attack.}
We observe a definite change in the strategy used by the C2 agent under an attacker's influence. Over a number of rollouts in either scenario, we frequently observe artifacts like erratic troop movements where the BlueForce troops keep oscillating about a single position for certain number of timesteps, straying off course of the aviation units which are crucial for winning the TigerClaw scenario, etc. Similarly in the NTC scenario, the attacks cause the BlueForce to retreat towards the bottom of the map, away from the advancing RedForce as opposed to following the original strategy -- to aggressively pursue and eliminate the RedForce units.

\begin{figure}[!ht]
    \centering
    \includegraphics[width=0.8\linewidth]{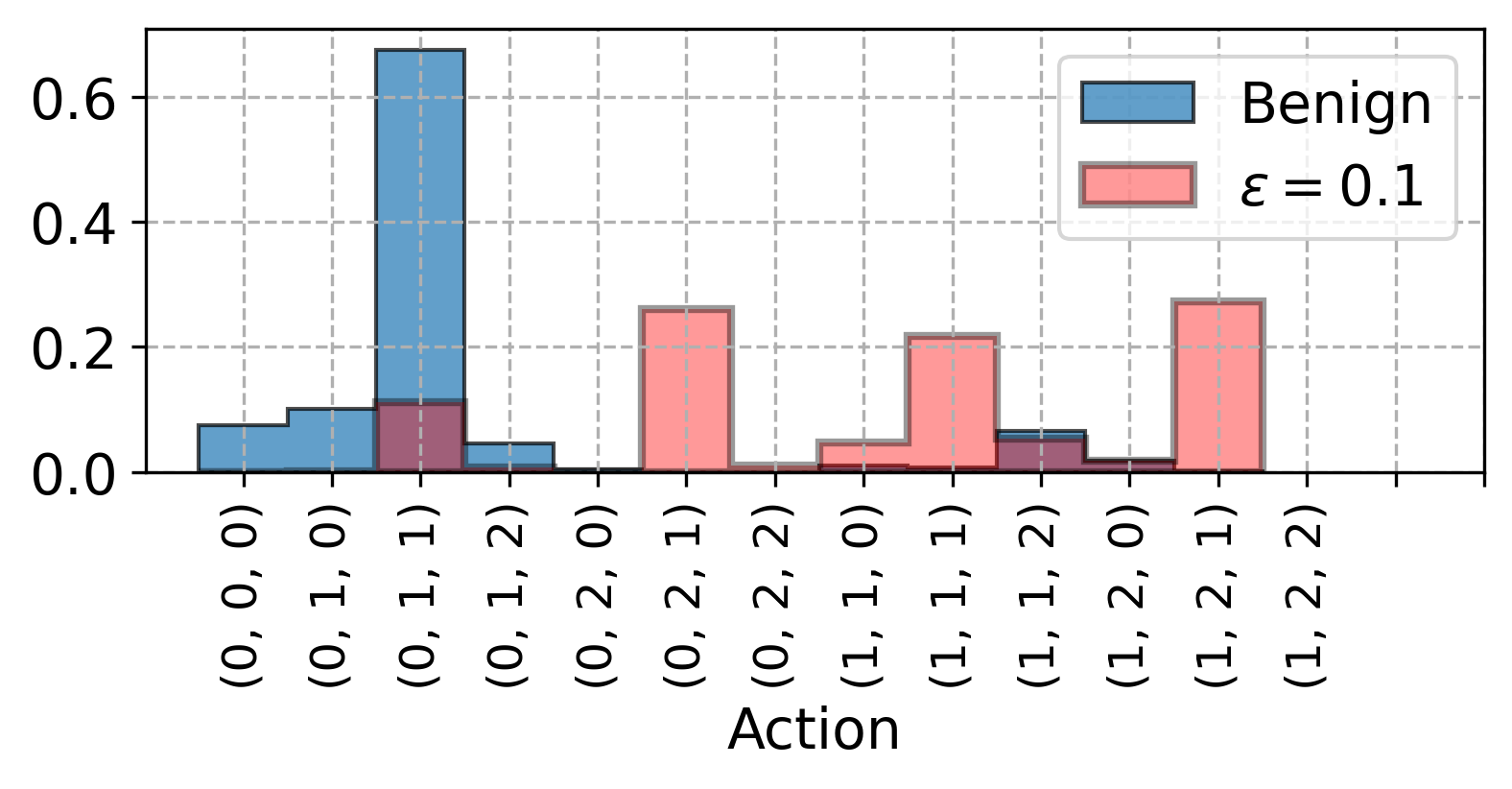}
    \caption{Sampled actions of a PPO agent over 5 TigerClaw rollouts. We can observe a shift in the distribution in the presence of an attacker.}
    \label{fig:action-dist-comp}
\end{figure}

To better understand this shift in behavior, we compare actions taken by the agent in a benign environment to those taken by the same agent when an adversary is present.
That is, at each timestep we compute the actions that \textit{would be taken} by the agent if the observations were maliciously modified and compare these ``subverted actions" to the actions that are actually taken by the agent at that timestep. 
We do so by perturbing the observations using our attack (with $\varepsilon =0.1$) and using the agent to predict the next action. The normalized action frequencies are plotted in Figure~\ref{fig:action-dist-comp} for a PPO/TigerClaw agent over 5 episodes. A clear difference is observed in the actions taken in the two cases. Actions $(0,2,1), (1,1,1), (1,2,1)$ are sampled frequently when perturbed observations are presented to the agent. These actions misdirect the BlueForce as they target areas on the map that do not contain any RedForce troops.

\begin{figure}[!ht]
    \centering \includegraphics[width=0.8\linewidth]{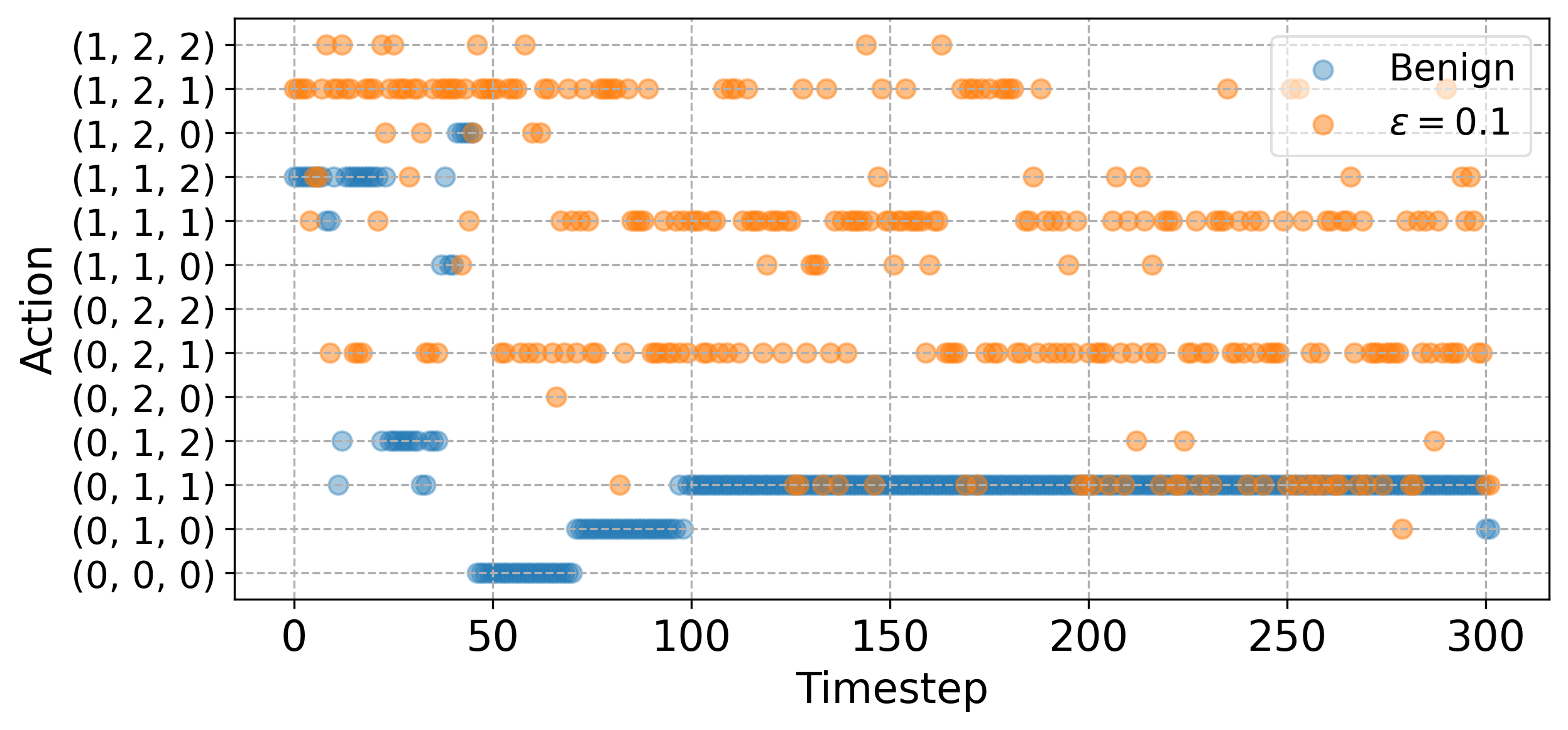}
    \caption{Actions taken by the BlueForce during a TigerClaw rollout in a benign setting compared to those taken in the presence of an inference time (FGSM) attacker.}
    \label{fig:action-comp}
\end{figure}

In order to examine the BlueForce movements more thoroughly, we select one episode and for each timestep plot the subverted actions and the actual actions taken (Figure~\ref{fig:action-comp}) by the agent. A rollout in a benign environment sees the BlueForce enter conflict and destroy most of the RedForce troops in the first 100 timesteps. This corresponds to the actions $(1,1,2), (1,1,1), (1,1,0)$ which are taken frequently within the first 50 timesteps. The $(0,1,1)$ actions correspond to \texttt{NO\_OP}s that are taken after exiting conflict when most of the RedForce are killed. In the presence of an attacker, however, in the first 50 timesteps the perturbations cause the C2 agent to sample actions like $(1,2,1)$ with high frequency. As a result, the BlueForce are misdirected and end up being killed by the RedForce. 

\begin{figure}[!ht]
    \centering
    \captionsetup[subfigure]{justification=centering}

    \begin{subfigure}{0.45\linewidth}
        \includegraphics[width=\linewidth]{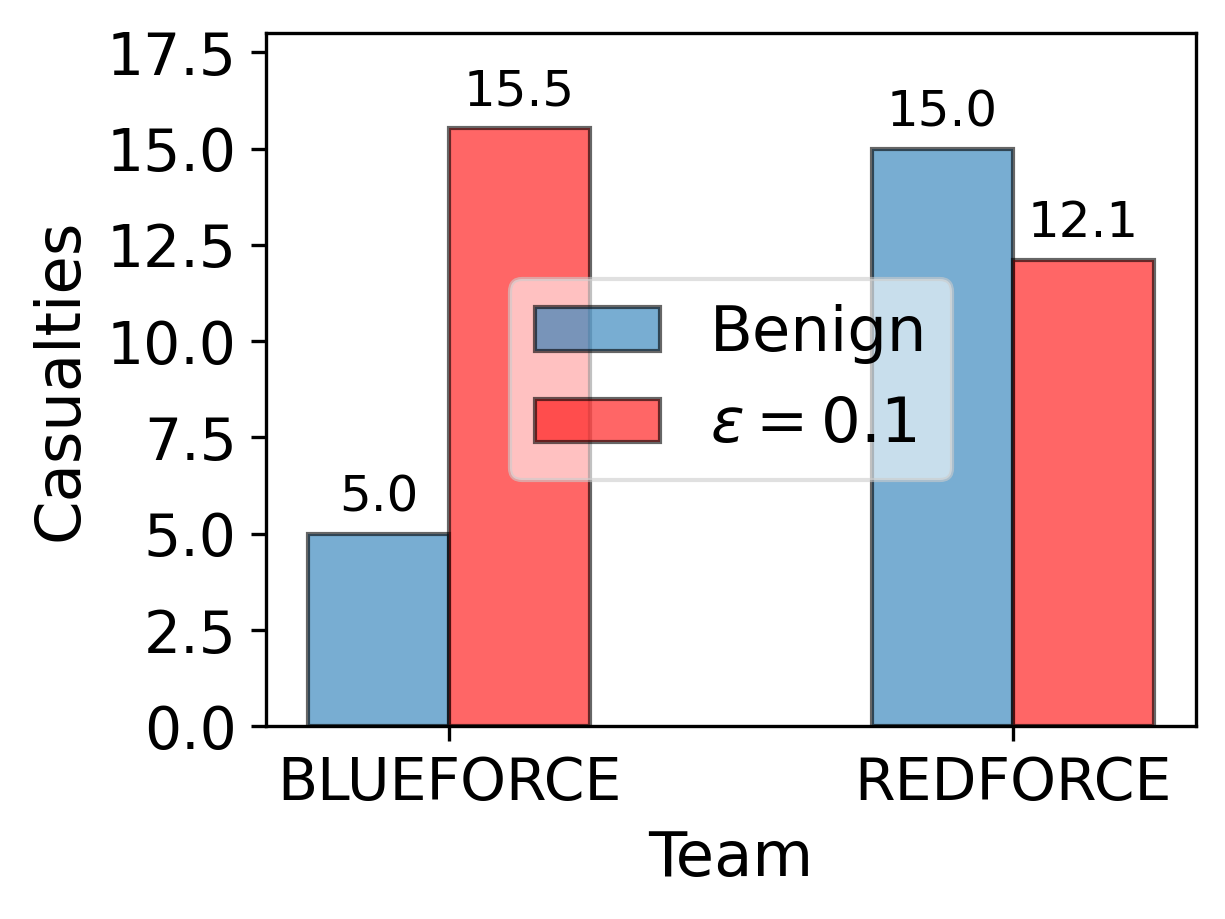}
        \caption{Casualty Comparison}
        \label{fig:metric-comp-1}
    \end{subfigure}
    \begin{subfigure}{0.45\linewidth}
        \includegraphics[width=\linewidth]{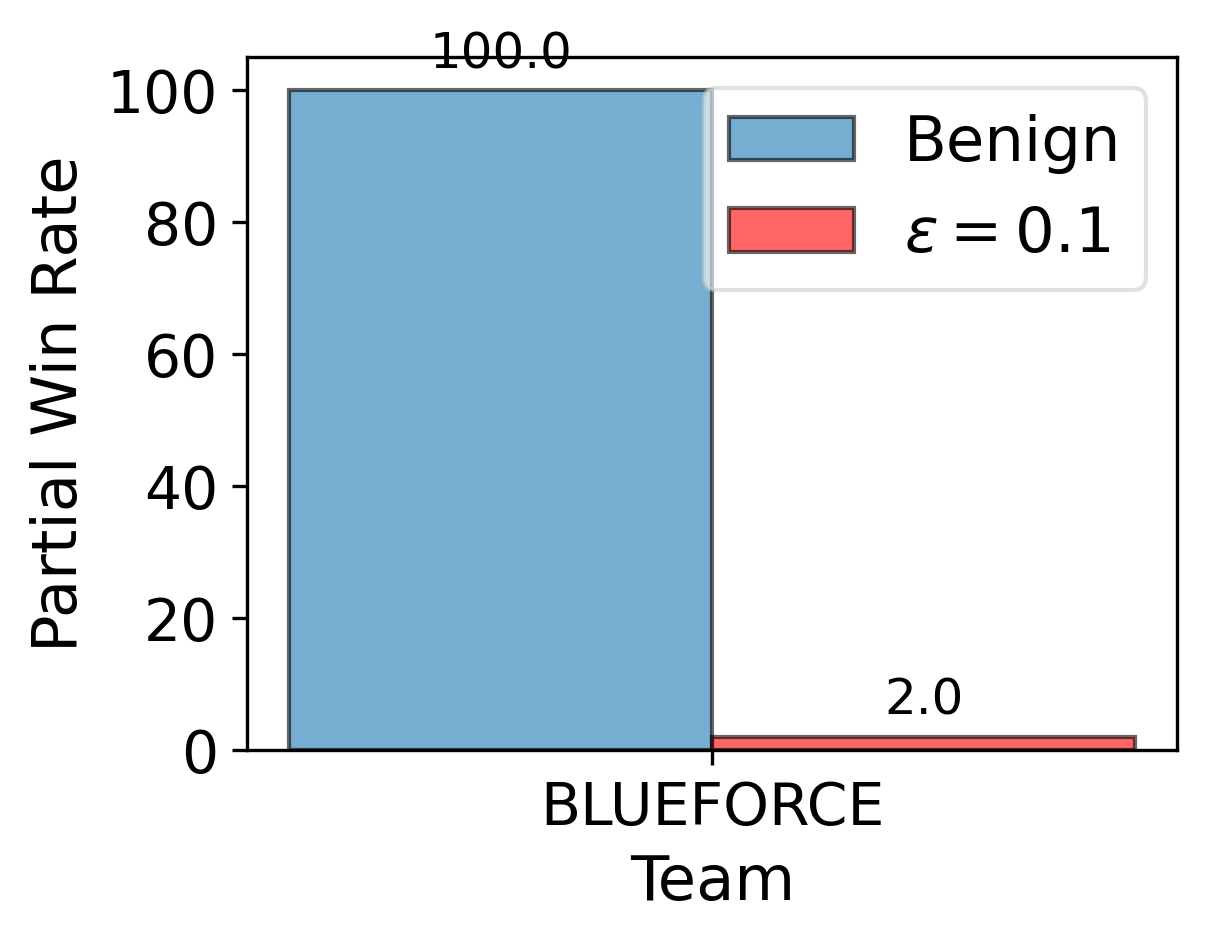}
        \caption{BlueForce Partial Win Rate}
        \label{fig:metric-comp-2}
    \end{subfigure}
    \begin{subfigure}{0.45\linewidth}
        \includegraphics[width=\linewidth]{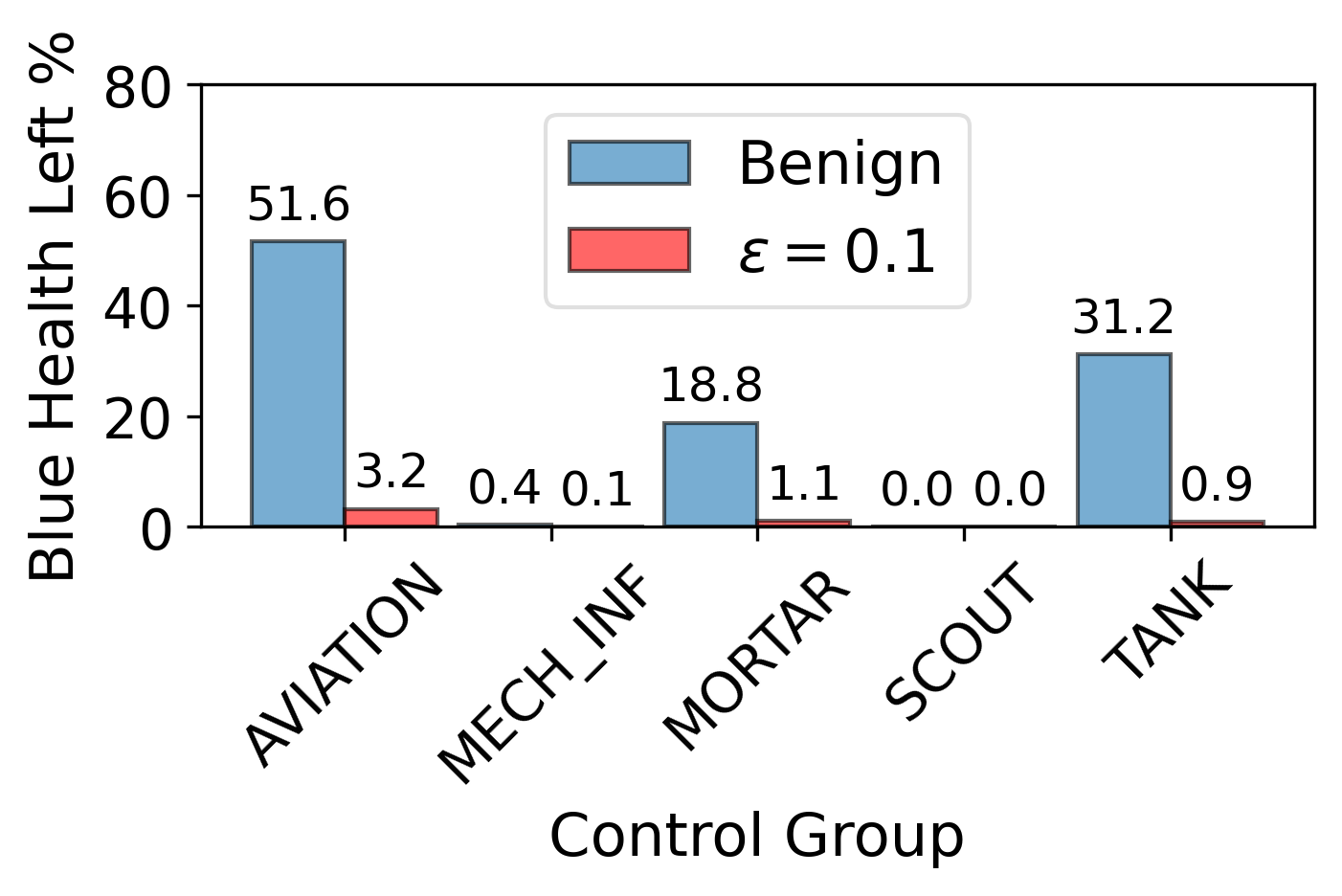}
        \caption{BlueForce Health Left}
        \label{fig:metric-comp-3}
    \end{subfigure}
    \begin{subfigure}{0.45\linewidth}
        \includegraphics[width=\linewidth]{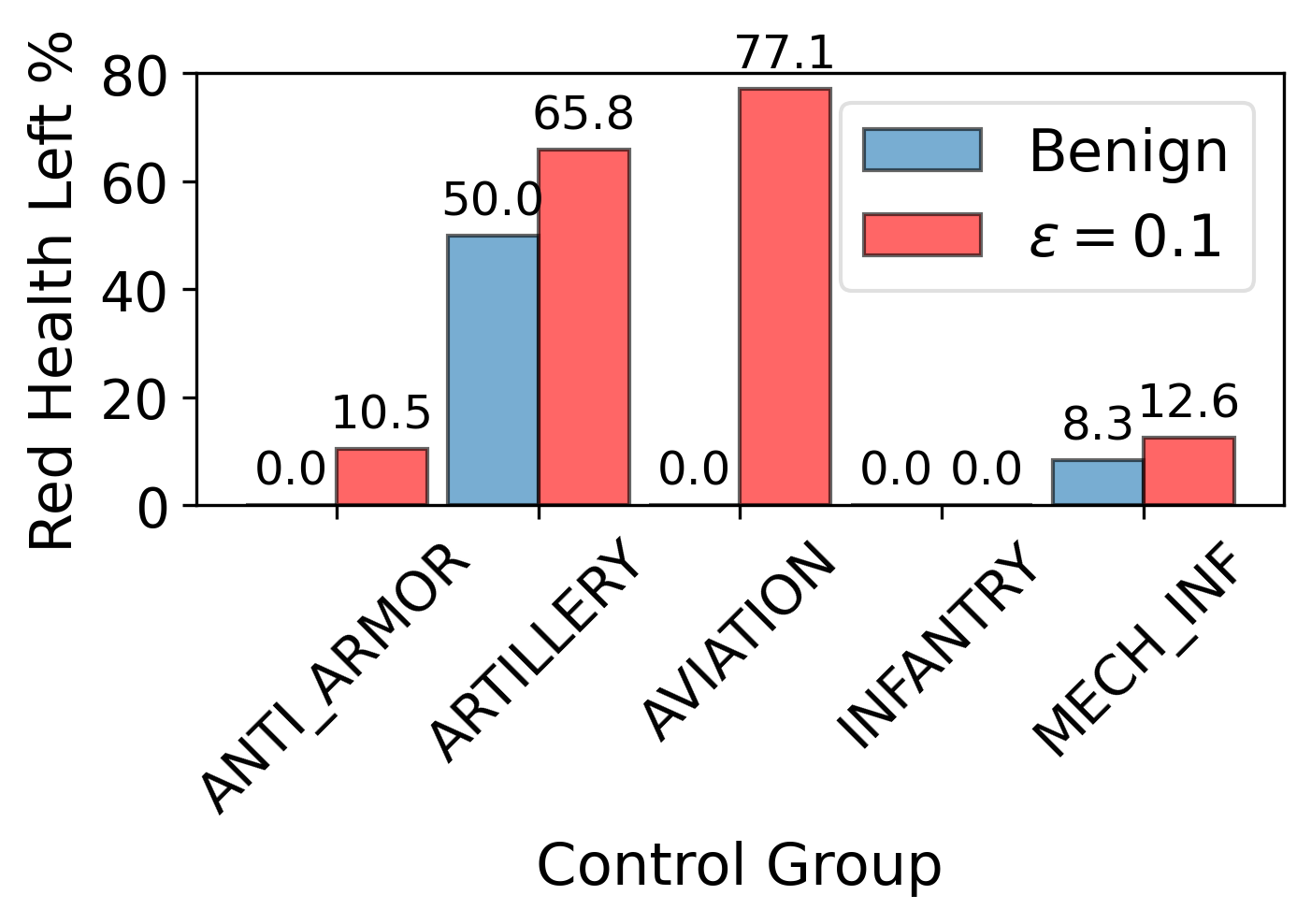}
        \caption{RedForce Health Left}
        \label{fig:metric-comp-4}
    \end{subfigure}
    \caption{Comparing the impact of inference time attack on additional game metrics for a PPO agent in TigerClaw. The results are aggregated over $100$ rollouts.}
    \label{fig:metric-comp}
\end{figure}

This is also reflected in the casualty metrics of both teams in Figure~\ref{fig:metric-comp-1}. There are greater BlueForce casualties and a fewer RedForce casualties in the presence of an $\varepsilon =0.1$ attacker. To get an idea of the impact of each unit on their respective team, we look at the health remaining percentage of each control group at the end of an episode. Figures~\ref{fig:metric-comp-3} and~\ref{fig:metric-comp-4} shows the health statistics for the BlueForce and RedForce respectively, aggregated over $100$ episodes. We can see in Figure~\ref{fig:metric-comp-3} that the health remaining of the \textit{aviation} units drops significantly to $2\%$ in the presence of an attacker. Rollouts show that these units are critical to winning the TigerClaw scenario for the BlueForce. On the other hand the RedForce \textit{aviation} units actually see an increase in remaining health ($77\%$). Other groups of the BlueForce also see a noticeable decrease in their health metrics.

\begin{figure}[!ht]
    \centering
    \captionsetup[subfigure]{justification=centering}
    % \hfill
    \begin{subfigure}{0.45\linewidth}
        \includegraphics[width=\linewidth]{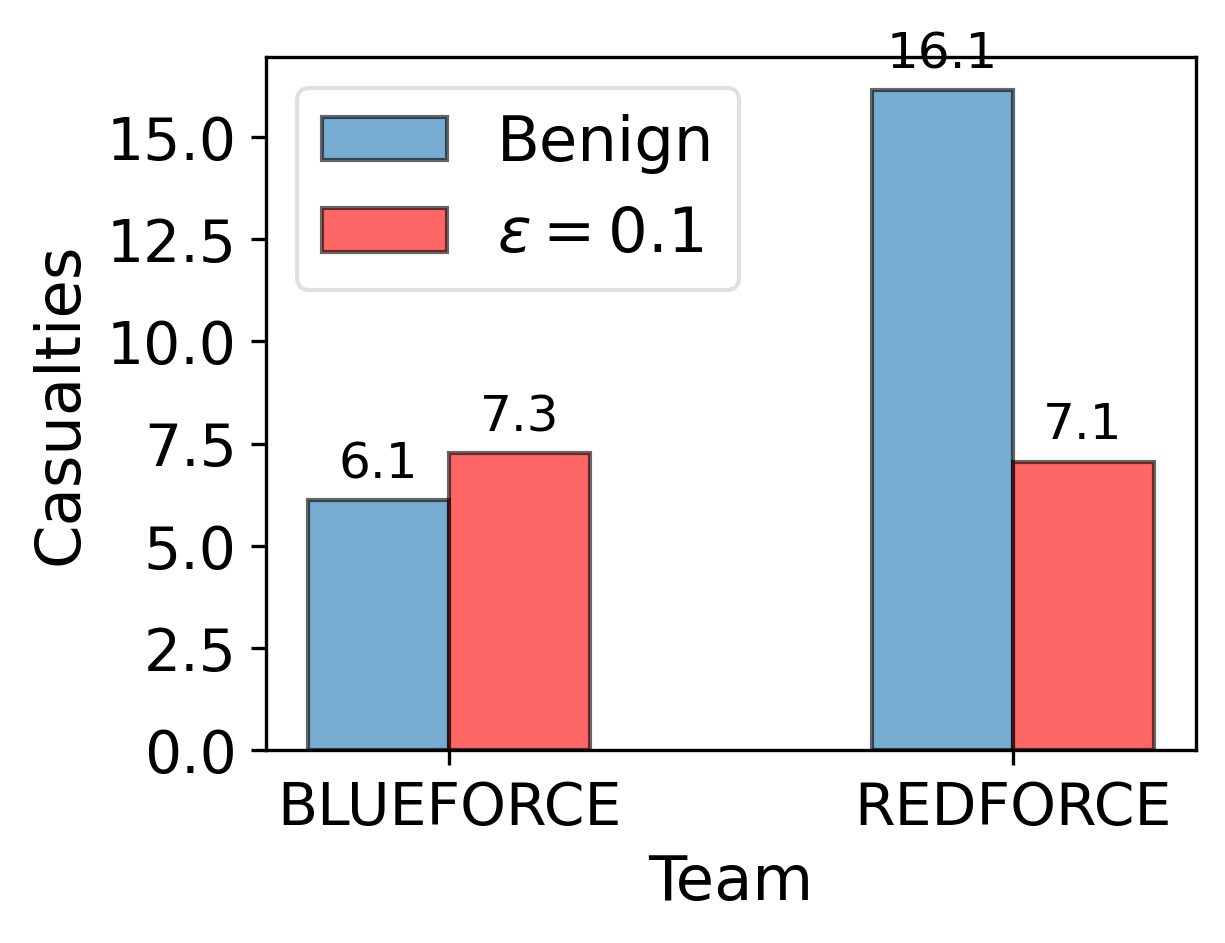}
        \caption{Casualty Comparison}
        \label{fig:ntc-metric-comp-1}
    \end{subfigure}
    \begin{subfigure}{0.45\linewidth}
        \includegraphics[width=\linewidth]{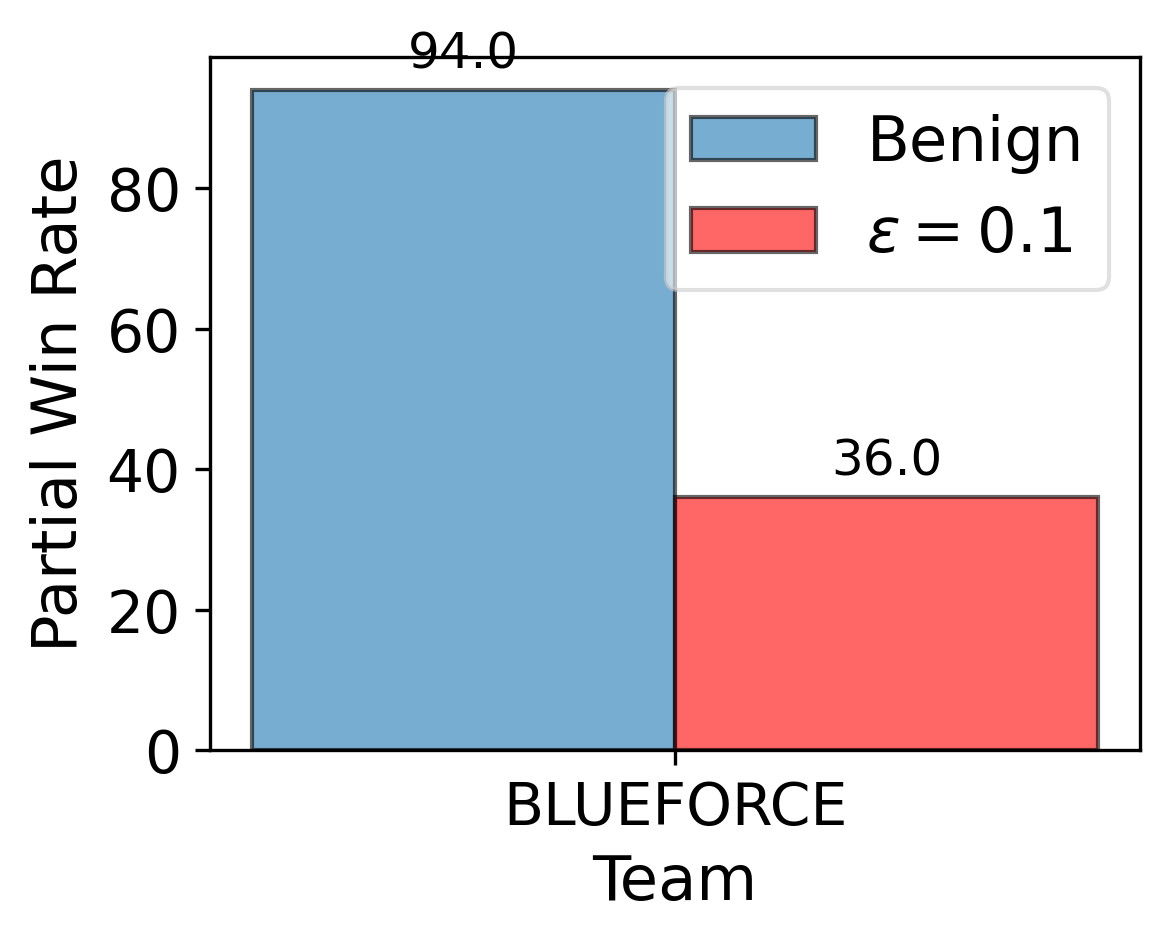}
        \caption{BlueForce Partial Win Rate}
        \label{fig:ntc-metric-comp-2}
    \end{subfigure}
    \begin{subfigure}{0.45\linewidth}
        \includegraphics[width=\linewidth]{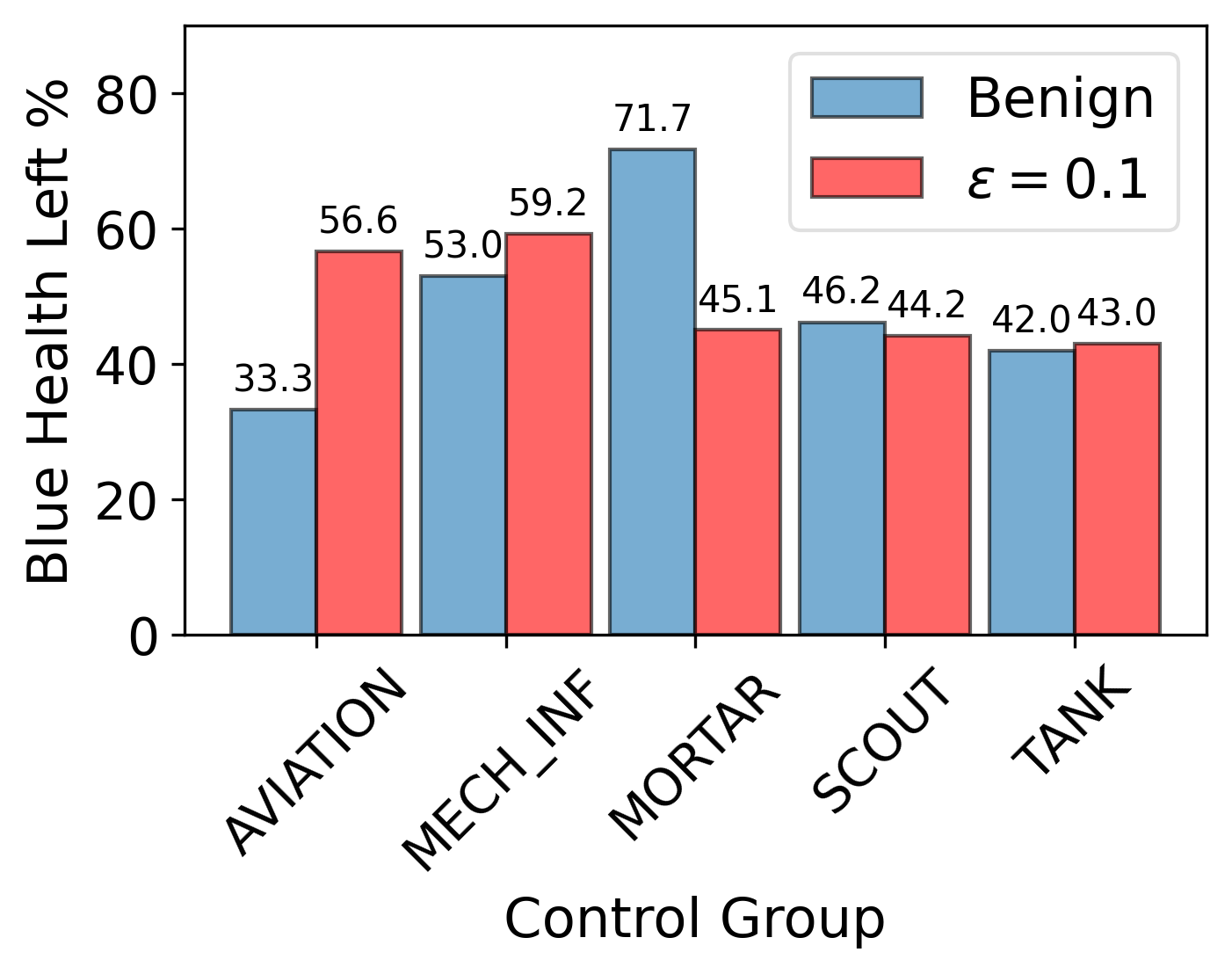}
        \caption{BlueForce Health Left}
        \label{fig:ntc-metric-comp-3}
    \end{subfigure}
    \begin{subfigure}{0.45\linewidth}
        \includegraphics[width=\linewidth]{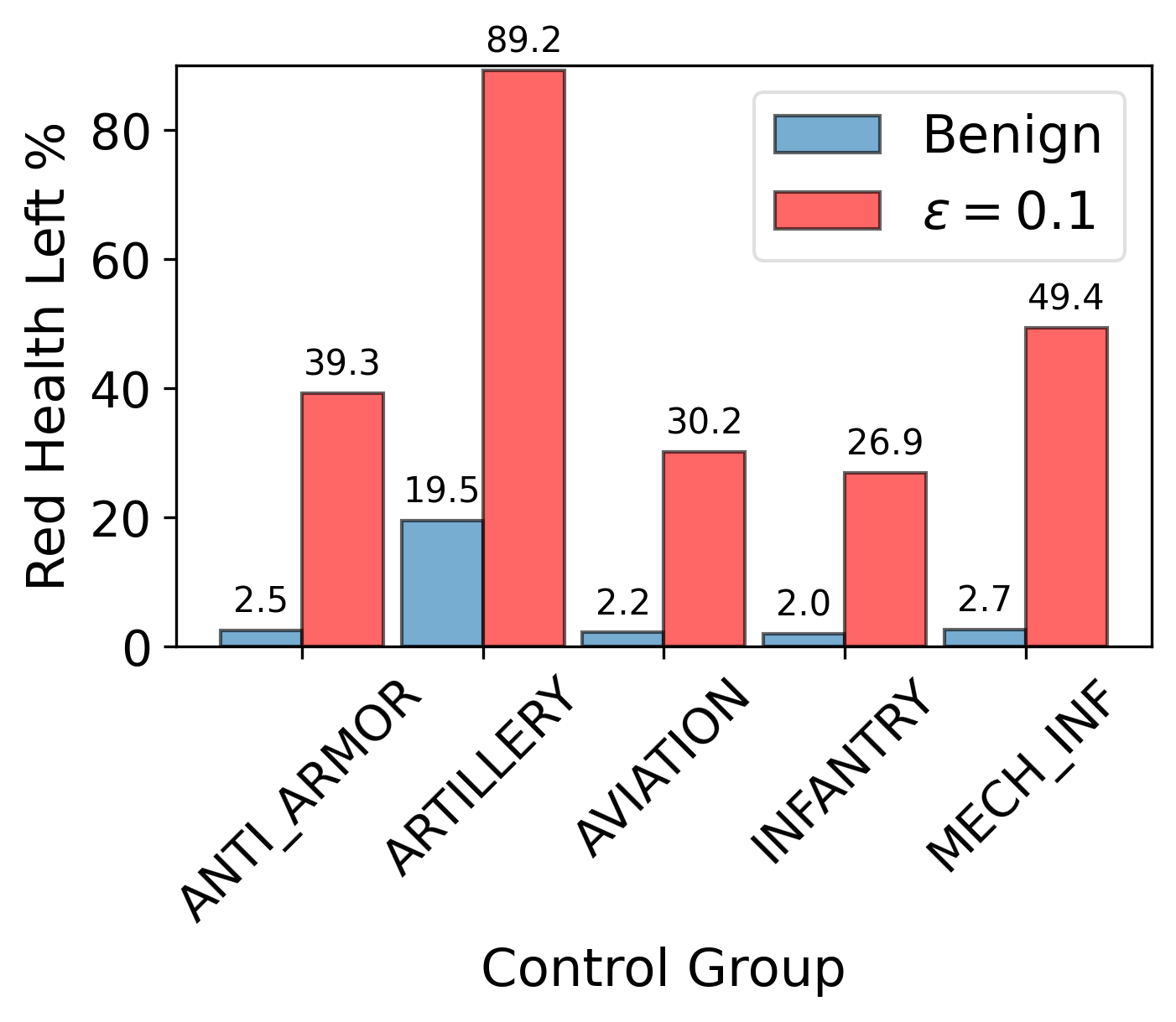}
        \caption{RedForce Health Left}
        \label{fig:ntc-metric-comp-4}
    \end{subfigure}
    \caption{Comparing the impact of inference time attack on additional game metrics for a PPO agent in NTC. The results are aggregated over $100$ rollouts.}
    \label{fig:ntc-metric-comp}
\end{figure}

Similarly in the NTC scenario, the health remaining for the RedForce increase greatly on attack (Figure~\ref{fig:ntc-metric-comp-4}), while the BlueForce health left only decreases marginally. Observing multiple rollouts reveals the attacks cause the BlueForce units to get misdirected  and as a result, they do not get into conflict with the RedForce. This is also supported by the casualty metrics for the two teams (Figure~\ref{fig:ntc-metric-comp-1}) with the BlueForce only having a slightly higher number of casualties when compared to the benign case.

Finally, the attack causes a sharp drop in the number of games that end with higher health levels for the BlueForce than the RedForce \textit{(termed a partial win)}. This can be seen in Figures~\ref{fig:metric-comp-2} and~\ref{fig:ntc-metric-comp-2}. 

\subsubsection{Strength \& Reliability of the Attack. }
For small perturbations, the attack is not very reliable as evidenced by the large dispersion in attained reward. 
High variance can be observed in Figures~\ref{fig:inf-PPO-1}, \ref{fig:inf-PPO-3}, \ref{fig:inf-A3C-1}, and~\ref{fig:inf-A3C-3} where we plot the (EMA smoothed) episode reward over $100$ episodes. This can be explained partly by the stochastic nature of action sampling and the reward structure associated with the map. This dispersion is reduced when increasing $\varepsilon$, leading to greater reliability but trading off the secrecy of the attack. A surprising instance is that of an A3C/TigerClaw agent where the attack remains ineffective even for larger perturbations. We analyze this in more detail below.\\
\noindent
\textbf{Attacking the A3C/TigerClaw agent. }
As can be seen in Figure~\ref{fig:inf-A3C-2}, the attack appears to fail on the C2 agent trained using A3C on the TigerClaw map. Even for large perturbations ($\varepsilon > 0.1$), the attack fails to degrade the agent's reward as extensively as the other cases. Interestingly, we also note that even in a benign environment, the A3C/TigerClaw agent achieves a significantly lower reward than the PPO/TigerClaw agent.

First, to rule out the training algorithm as a potential factor and for a fair comparison, we partially train a PPO agent on the TigerClaw scenario until it achieves a similar reward to the A3C agent and study the rewards achieved by this agent in the presence of an attacker.

\begin{figure}[!ht]
    \centering
    \begin{subfigure}{0.48\linewidth}
        \includegraphics[width=\linewidth]{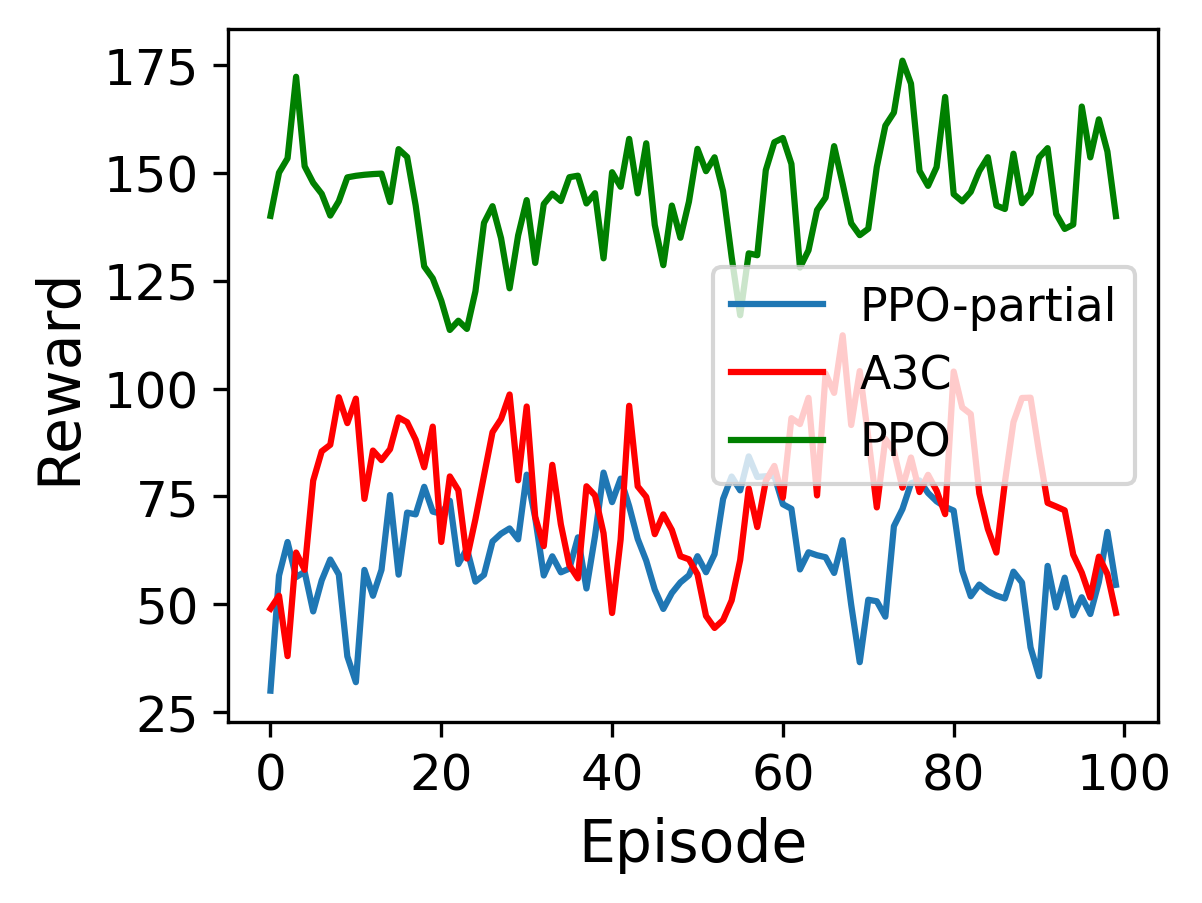}
        \caption{Episode rewards of the agent under benign conditions.}
        \label{fig:benign-comp-1}
    \end{subfigure}
    \begin{subfigure}{0.48\linewidth}
        \includegraphics[width=\linewidth]{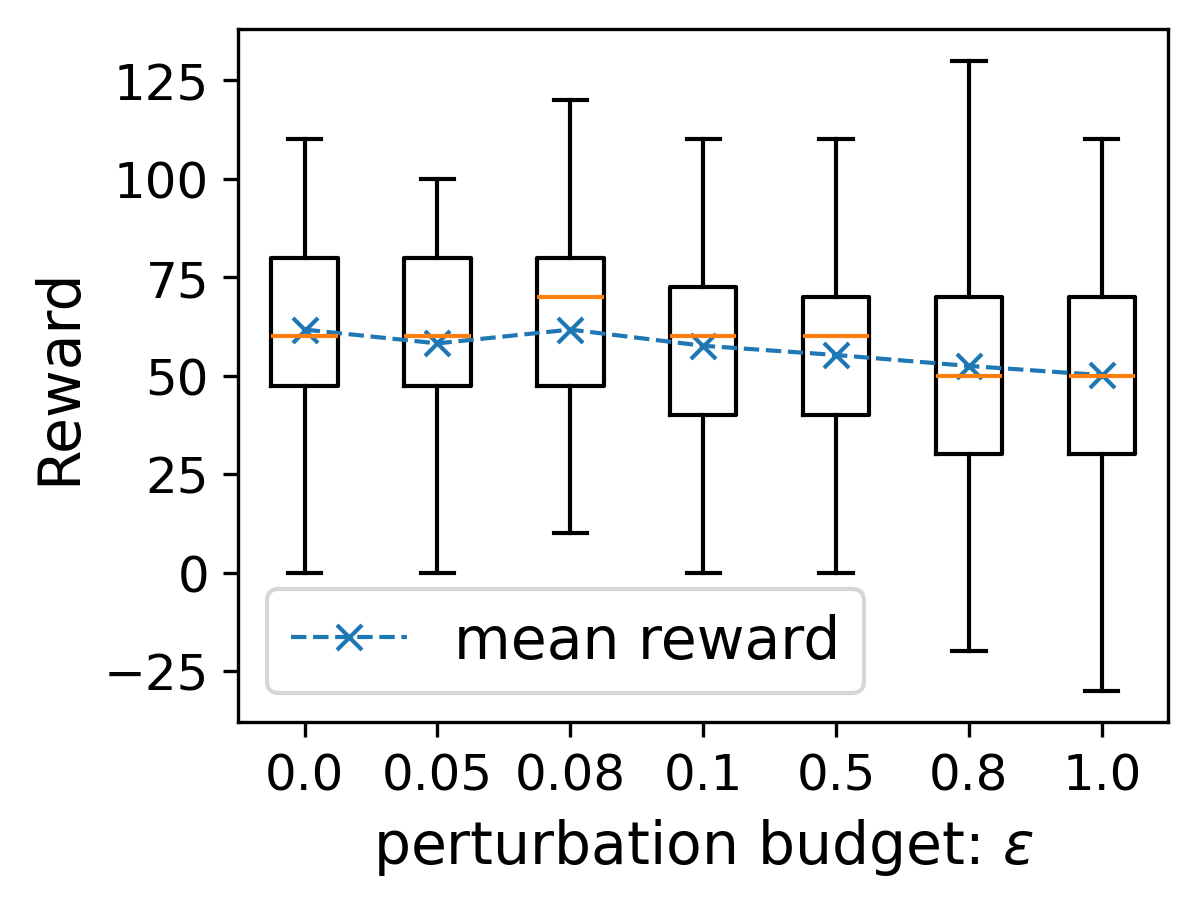}
        \caption{PPO-partial/TigerClaw: Reward trend w.r.t $\varepsilon$}
        \label{fig:benign-comp-2}
    \end{subfigure}
    \caption{Comparing the effect of the inference time attack on the A3C agent with the partially trained PPO agent (PPO-partial). This agent was trained for 1M timesteps. }
    \label{fig:benign-comp}
\end{figure}

Our observations are shown in Figure~\ref{fig:benign-comp}. In Figure~\ref{fig:benign-comp-1} we see that the partially trained PPO agent \textit{\textbf{(PPO--partial) }}gets similar rewards as the A3C agent over 100 rollouts. 
Notably, its reward trend when attacked (Figure~\ref{fig:benign-comp-2}) is comparable to that of the A3C agent (Figure~\ref{fig:inf-A3C-2}). This empirically shows that the attack is largely unaffected by the training algorithm.
Coupled with the previous observation, we hypothesize that the effectiveness of the attack is correlated to the quality of the trained agent. 

\begin{figure}[!ht]
    \centering
    \captionsetup[subfigure]{justification=centering}
    \begin{subfigure}{0.49\linewidth}
        \includegraphics[width=\linewidth]{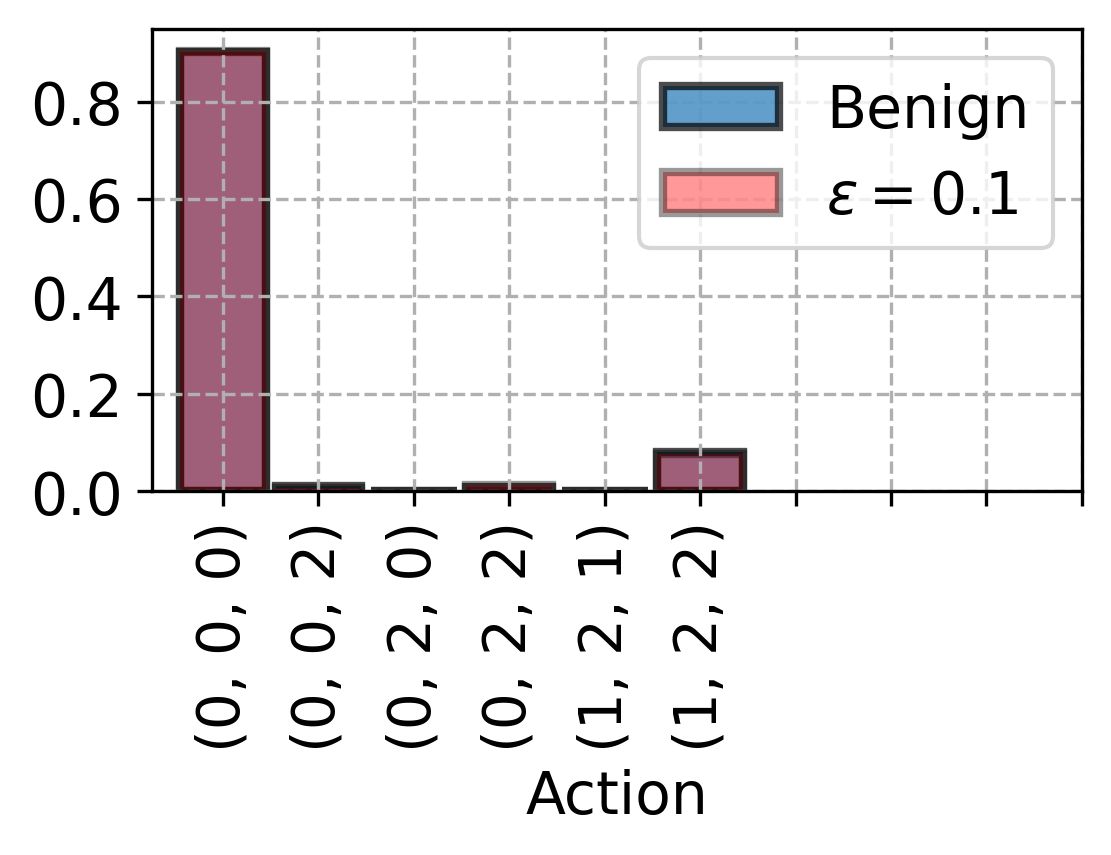}
        \caption{A3C/TigerClaw}
        \label{fig:ap1}
    \end{subfigure}
    \begin{subfigure}{0.49\linewidth}
        \includegraphics[width=\linewidth]{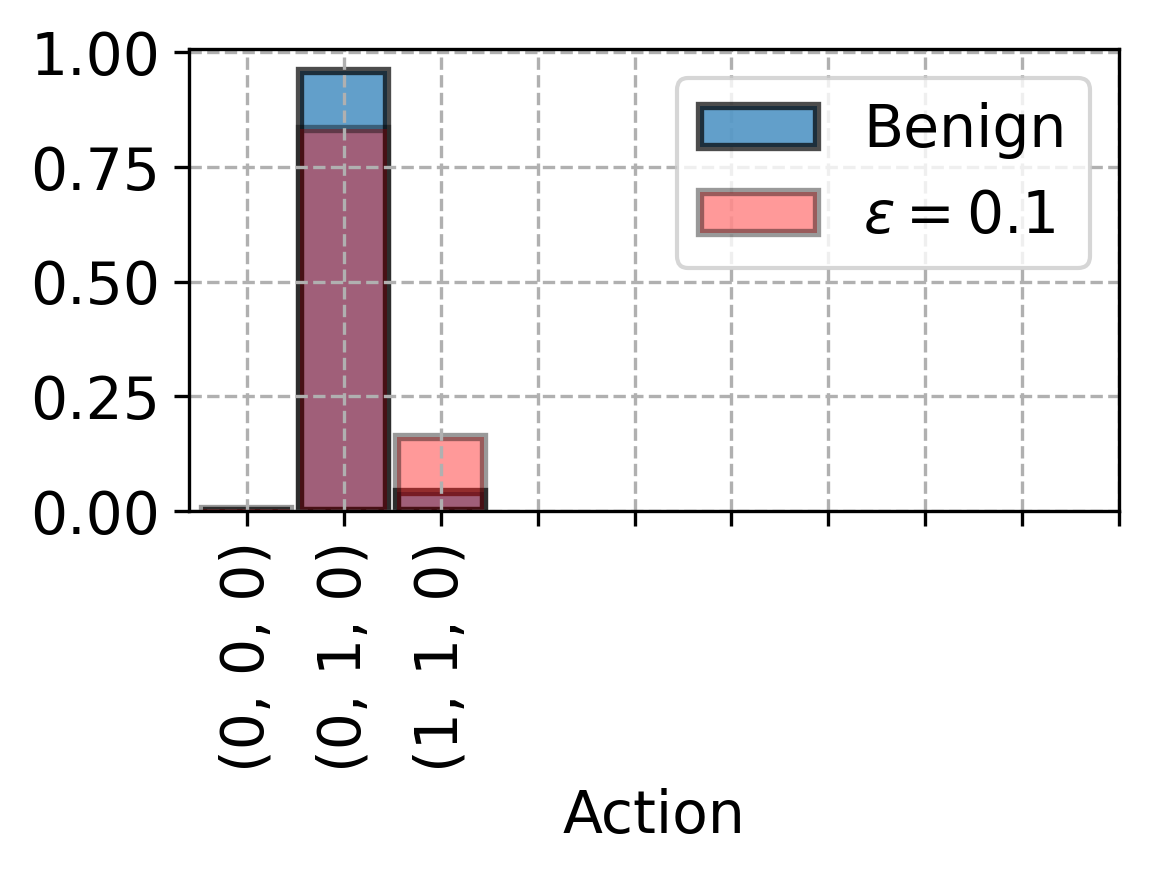}
        \caption{PPO-Partial/TigerClaw}
        \label{fig:ap5}
    \end{subfigure}
    \caption{Action distribution shifts over 5 episodes for A3C/TigerClaw and PPO-Partial/TigerClaw}
    \label{fig:ap}
\end{figure}

To test this hypothesis in greater detail we turn to the frequencies of actions taken by both agents over multiple rollouts.
Similar to the analysis in previous sections, to visualize the shift induced by the attack we plot the subverted and actual ($\varepsilon = 0.1$) actions sampled over 5 episodes in Figure~\ref{fig:ap} for both agents.
Surprisingly, we observe an almost complete overlap of the two action plots for the A3C agent (Figure~\ref{fig:ap1}) and only a marginal difference for the PPO-partial agent. This indicates that the attack is in most cases incapable of flipping the actions sampled by the agent.

\begin{figure}[!ht]
    \centering
    \includegraphics[width=0.9\linewidth]{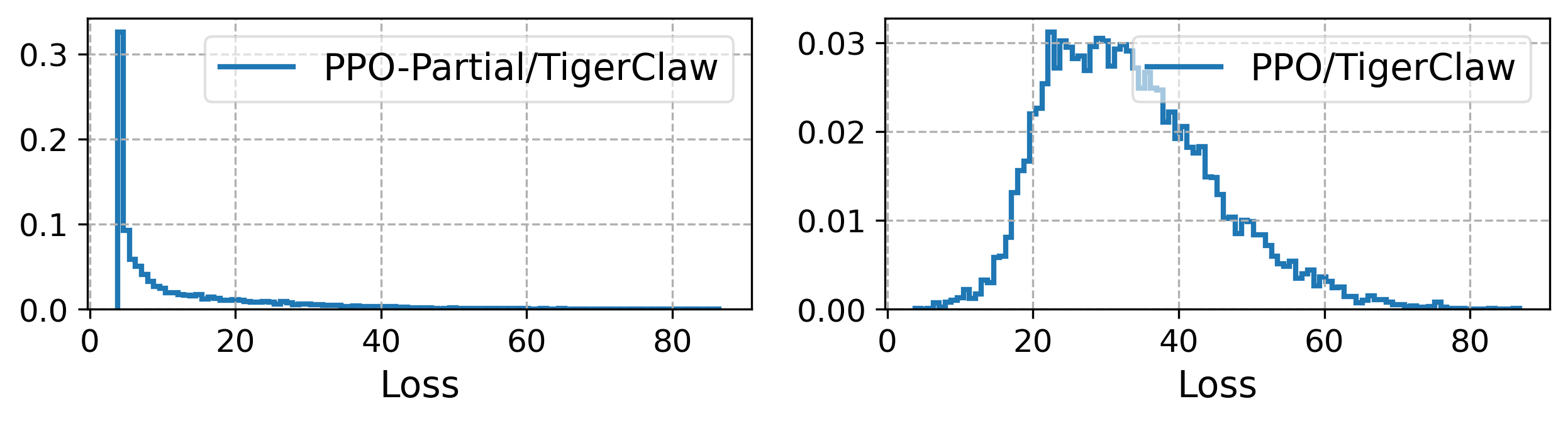}
    \includegraphics[width=0.9\linewidth]{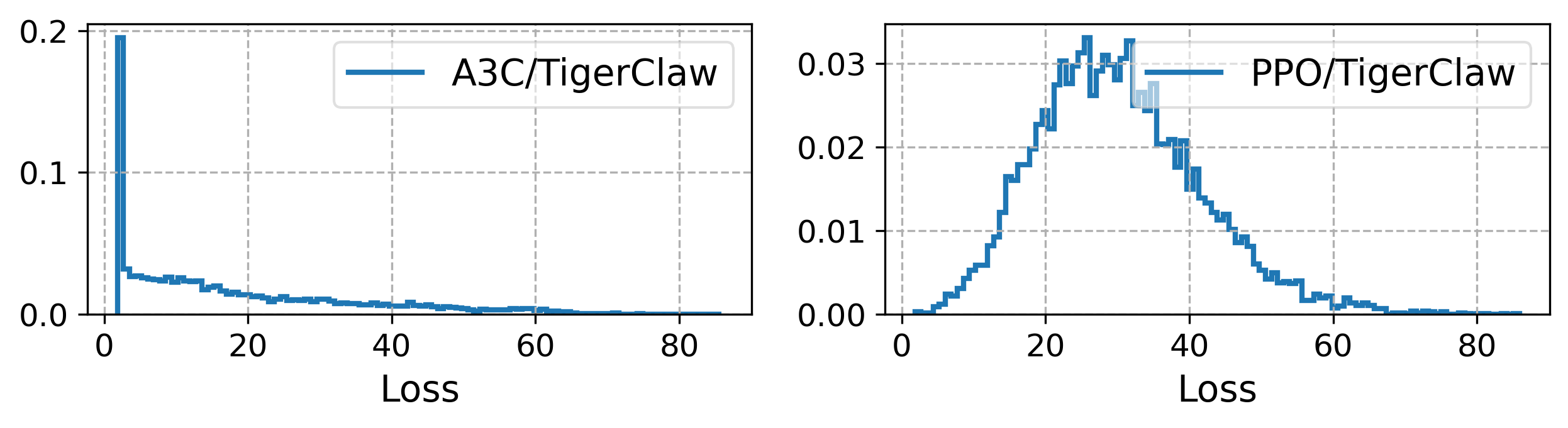}
    \caption{Loss Value plots for the agent policy networks. The $y$ axis represent the normalized frequencies ($10^4$ trials) for the loss value on the $x$ axis. }
    \label{fig:loss-landscape}
\end{figure}

\begin{figure*}[!ht]
    \centering
    \begin{subfigure}{0.24\linewidth}
        \includegraphics[width=\linewidth]{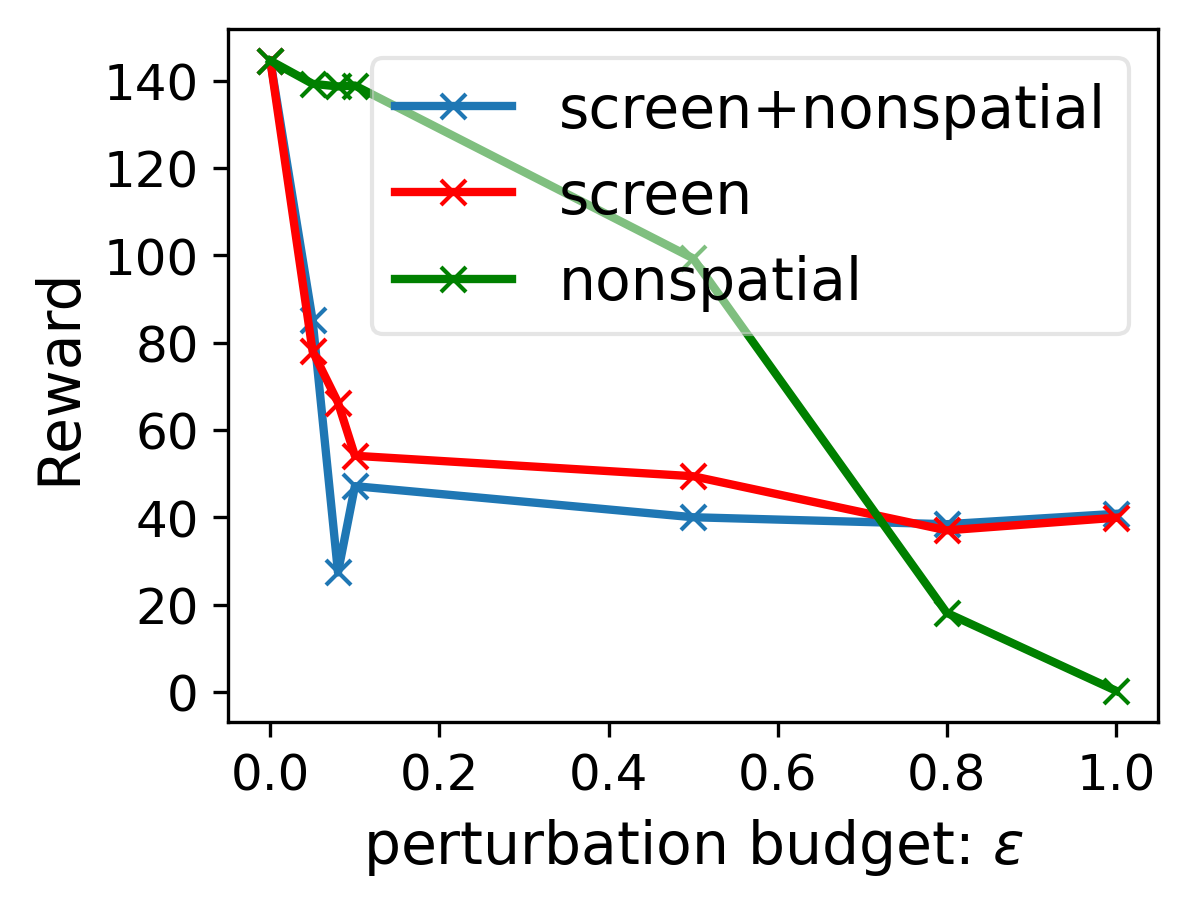}
        \caption{PPO/TigerClaw Agent mean reward comparison.}
        \label{fig:componentwise-eval-1}
    \end{subfigure}
    \begin{subfigure}{0.24\linewidth}
        \includegraphics[width=\linewidth]{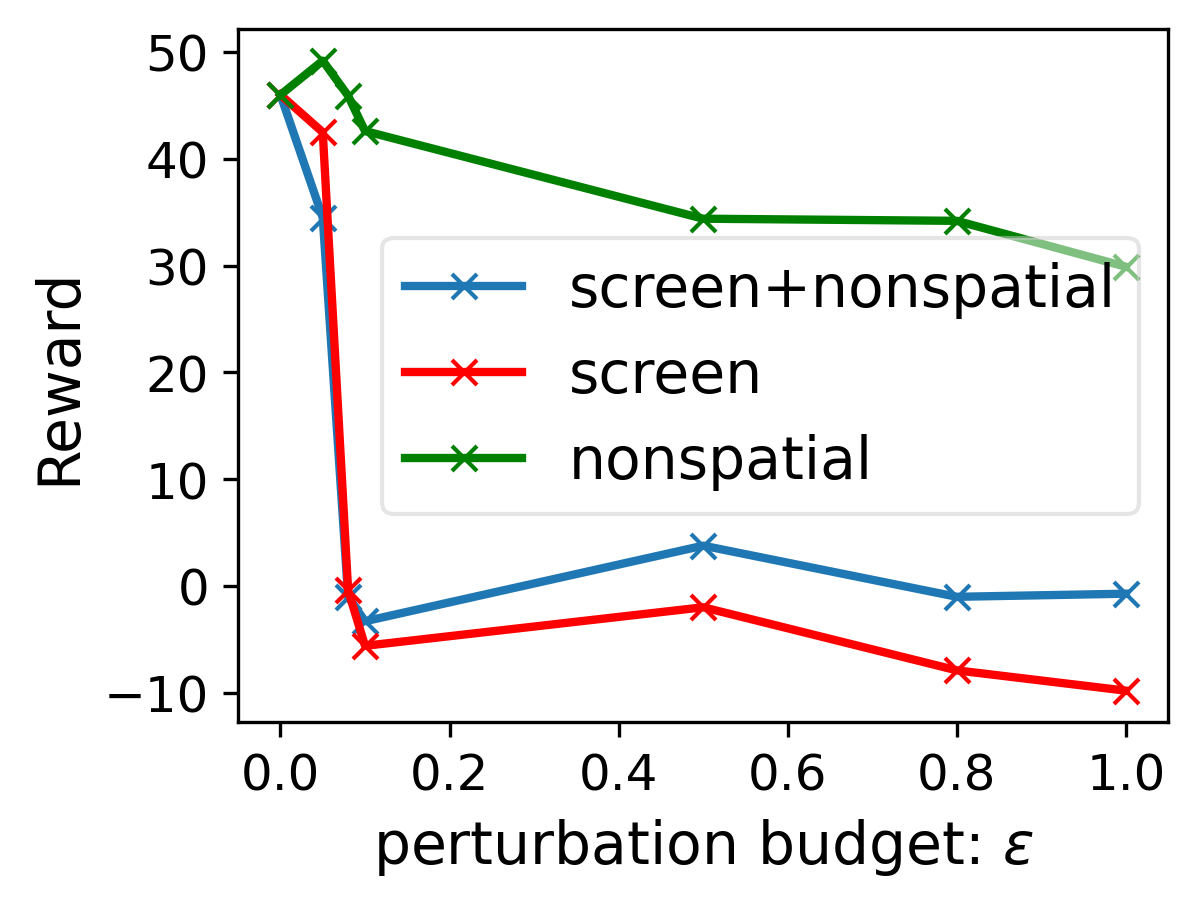}
        \caption{PPO/NTC Agent mean reward comparison.}
        \label{fig:componentwise-eval-2}
    \end{subfigure}
    \begin{subfigure}{0.24\linewidth}
        \includegraphics[width=\linewidth]{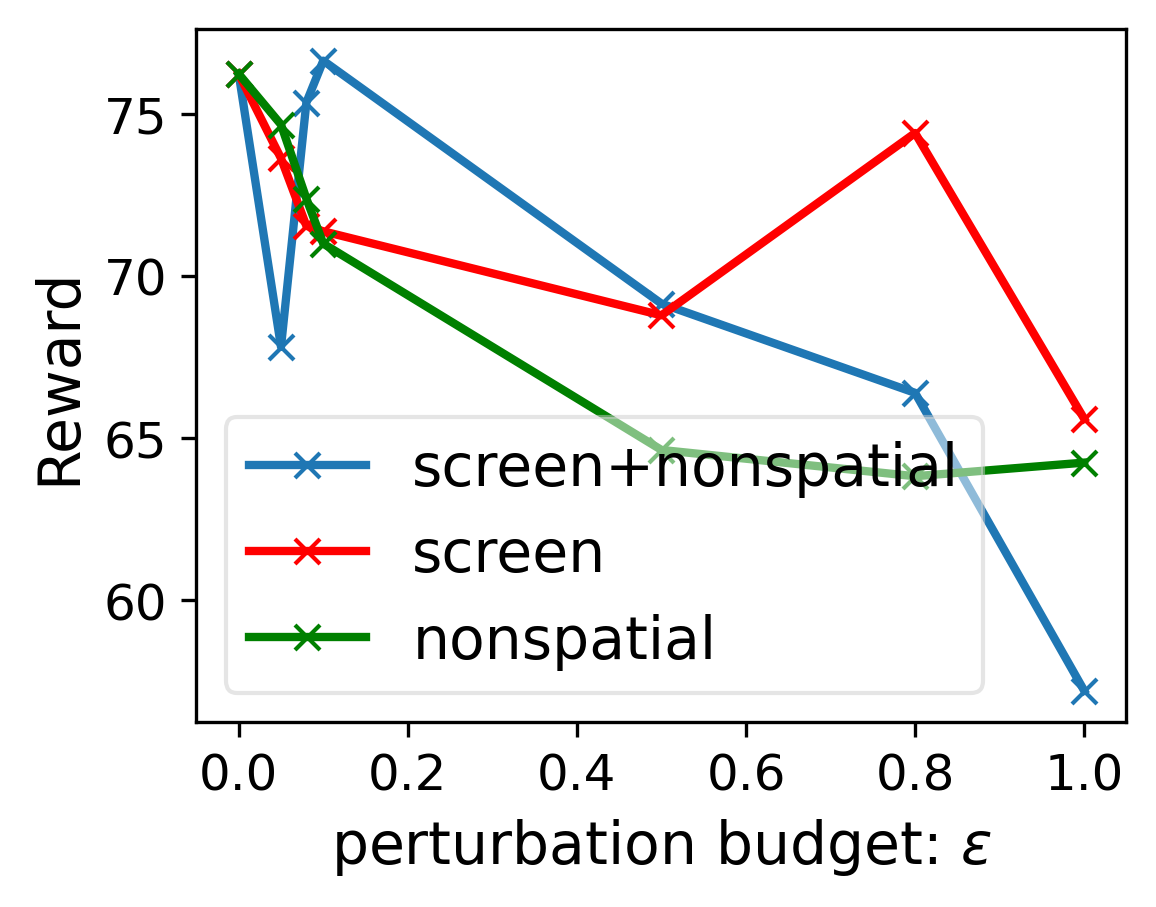}
        \caption{A3C/TigerClaw Agent mean reward comparison.}
        \label{fig:componentwise-eval-3}
    \end{subfigure}
    \begin{subfigure}{0.24\linewidth}
        \includegraphics[width=\linewidth]{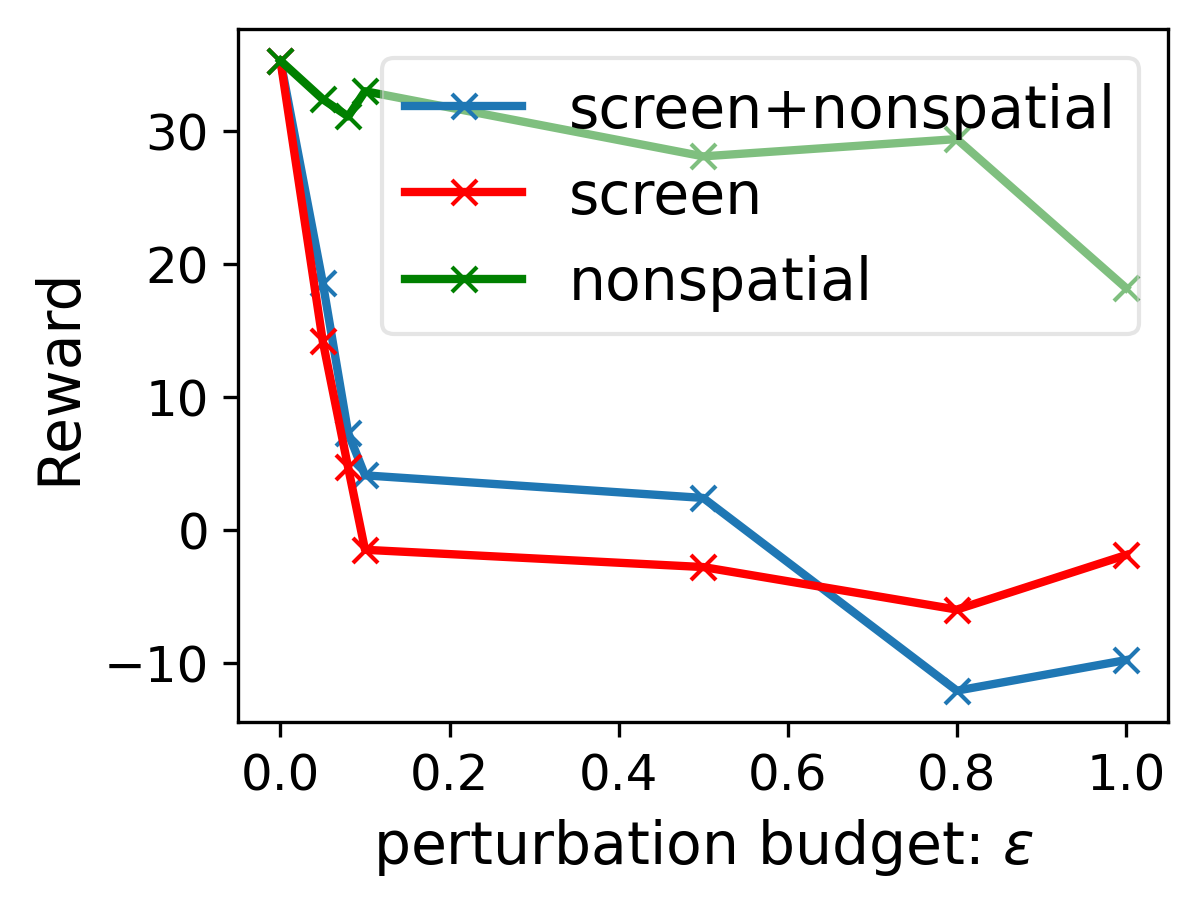}
        \caption{A3C/NTC Agent mean reward comparison.}
        \label{fig:componentwise-eval-4}
    \end{subfigure}
    \caption{Component wise impact of the input on mean episode reward.}
    \label{fig:componentwise-eval}
\end{figure*}

As the malicious perturbations are constructed using a loss gradient, we try to understand the loss landscape of the poorly trained policy networks to explain the attack's ineffectiveness. To do so, we plot the loss value for $10^{4}$ randomly sampled points in the  $l_\infty$-ball of $\varepsilon$ radius around a fixed observation $O_c$. The loss is computed as the component-wise sum of the CE loss between the predicted action distribution on that observation and a fixed ground-truth action distribution.
The ground truth action distribution 
is computed corresponding to the actual action taken by the agent on $O_c$ (Section~\ref{sec:inf-time-attack}). The CE loss is calculated for each output component (action logit, x-value, y-value) and summed to get the final loss. Figure~\ref{fig:loss-landscape} shows the results for $\varepsilon = 0.1$ for both the A3C and PPO-Partial agents when compared to the PPO/TigerClaw agent as a baseline. Compared to the PPO agent, both the A3C and PPO-Partial agent's predictions are highly similar to the ground truth as evidenced by the highly frequent $\sim 0$ loss value. This corresponds to a \textit{flatness} in the prediction space of both (A3C and PPO-Partial) policy networks where inputs in neighbourhood result in the same prediction. Consequently this leads to greater robustness to injected noise -- benign and adversarial as small perturbations are not enough to significantly change the predicted action distributions. Investigating the relation between this perceived robustness and the quality of training is left as future work.

\subsubsection{Component-wise Impact of Input on Attack Results. }
In this section we analyze the impact of the screen and nonspatial components on the effectiveness of the attack. To do so, we perform the attack by restricting malicious perturbations to only the screen or nonspatial components respectively. We track the mean reward achieved by the C2 agent in each case and compare them to the baseline attack which perturbs both components. We record the rewards gained over 100 rollouts and show the results in Figure~\ref{fig:componentwise-eval}.

Figures~\ref{fig:componentwise-eval-1}, \ref{fig:componentwise-eval-2}, and~\ref{fig:componentwise-eval-4} indicate that for smaller perturbations levels ($\epsilon <0.5$), modifications to the screen component are primarily responsible for a successful attack. Further, we only see a minor degradation in the reward for small nonspatial-only perturbations in both scenarios. On the other hand, the drop in mean reward because of screen-only perturbations is similar to that in the baseline attack. Intuitively, this skewed importance makes sense as the screen component is a much larger input component ($256 \times 256$ versus $287$) 
when compared to the nonspatial component. The screen component further encodes important spatial information on the terrain and location of troops that has a significant impact on the policy network's prediction.

We also note the anomalous trends in Figure~\ref{fig:componentwise-eval-3} for the A3C/TigerClaw agents. It appears that the nonspatial component is more important to the policy network, however since this agent is poorly trained, we cannot conclusively reason about the resulting trends. 

\subsubsection{Adversarial Robustness of the Training Algorithm.}
\begin{figure}[!ht]
    \centering
    % \hfill
    \begin{subfigure}{0.48\linewidth}
        \includegraphics[width=\linewidth]{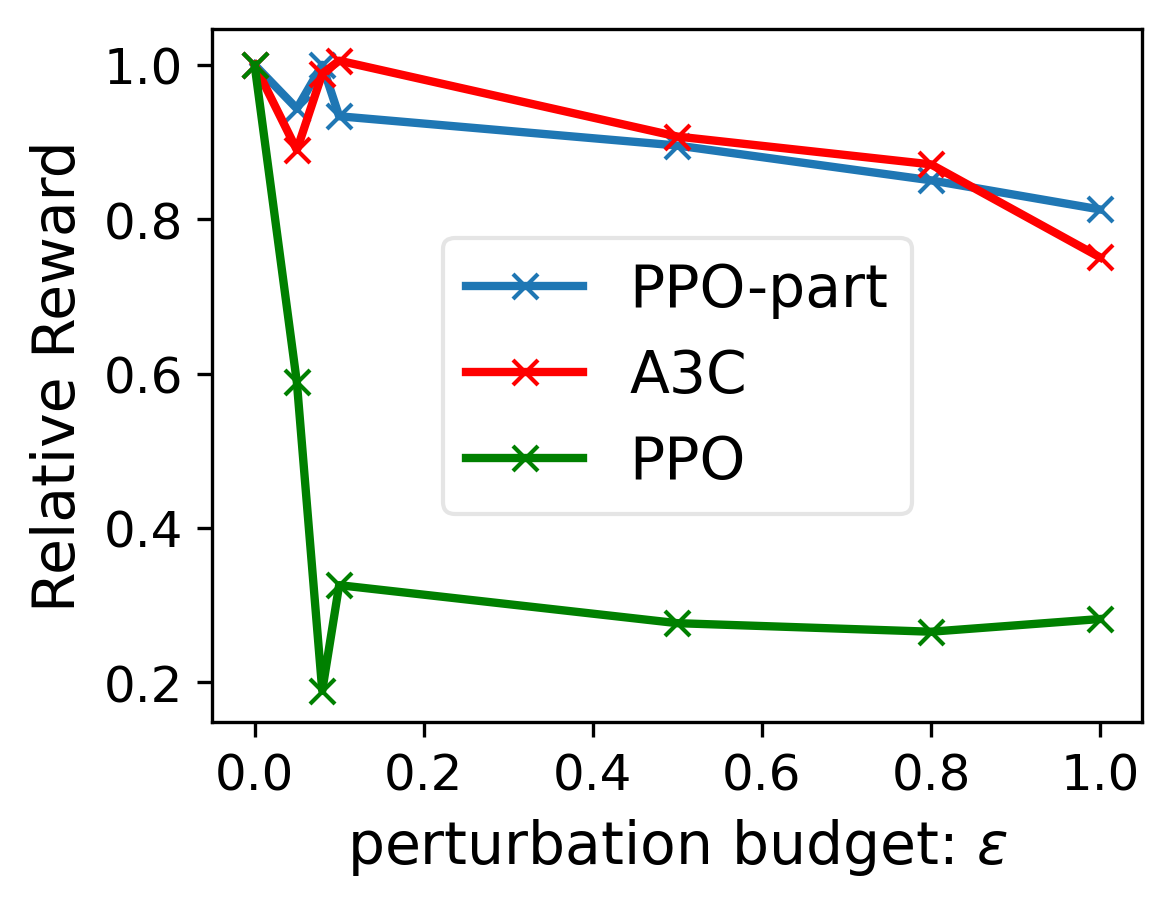}
        \caption{Relative Rewards of PPO vs A3C on TigerClaw}
        \label{fig:algo-comp-1}
    \end{subfigure}
    \begin{subfigure}{0.48\linewidth}
        \includegraphics[width=\linewidth]{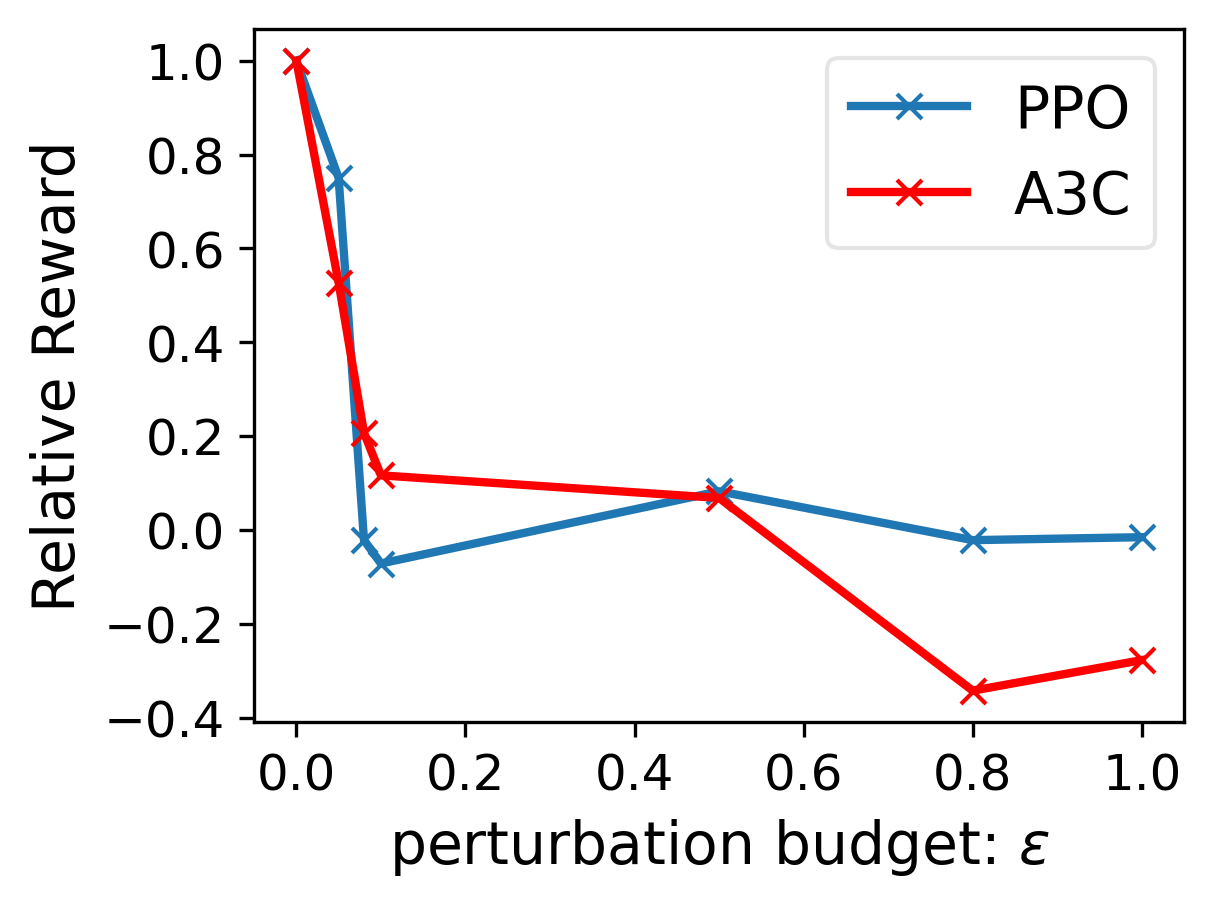}
        \caption{Relative Rewards of PPO vs A3C on NTC}
        \label{fig:algo-comp-2}
    \end{subfigure}
    \caption{Comparison of Algorithm Robustness: Mean Relative Reward with respect to the perturbation budget}
    \label{fig:algo-comp}
\end{figure}

To compare the robustness of an agent trained using PPO to one trained using A3C we compare the trend followed by the \textit{mean relative reward} $(R_r)$ defined as the ratio of the mean reward obtained by an agent under attack $(R_a)$ to the mean reward earned in a benign environment $(R_b)$. That is, $R_r = {R_a}/{R_b}$. A larger value of $R_r$ signifies more robustness.
Figure~\ref{fig:algo-comp} presents the results of our comparisons. In both cases, we see that the A3C agent seems to perform marginally better than the PPO agent for small perturbations and worse for higher perturbation levels.
As in the previous section, for a fair comparison we compare the partially trained PPO agent (PPO-partial) with the A3C agent (Figure~\ref{fig:algo-comp-1}) and see a similar decreasing trend for $R_r$. 

While we cannot conclusively comment on the robustness of A3C vs PPO, the robustness of the PPO-partial agent compared to a fully trained PPO agent makes for a compelling argument. This has been explored in more detail in the preceding section. It should be noted however that PPO does seem to be functionally more performant than A3C as evidenced by the higher rewards in both scenarios.
\section{Discussions and Future Work}
\label{sec:disc}

In this work we have focused on evaluating the robustness of RL agents used for C2 through the lens of an inference time attacker. Our investigations supported by existing literature reveal that even the latest RL training algorithms cannot be trusted to train agents that can be reliably deployed in unsafe environments. Our evaluations show that well-trained agents are highly sensitive to even minute perturbations in their input space and act suboptimally as a result. In an arena such as the battlefield where observations received can be prone to noise either benign or malicious, this raises serious questions on the use of such agents. 

This directly leads us to two avenues for future work. The first is developing explainable and controllable approaches to COA generation using RL training and developing robustness mechanisms for RL agents that work both during inference and training time. Such mechanisms can be a combination of adversarial perturbation detection and prevention mechanisms that can be deployed on top of pre-existing agents. The second avenue for future work is existing training algorithms can be augmented to train agents that are certifiably robust to malicious noise. In this vein, adversarial training offers a promising alternative to train robust agents. In this work, while we present a preliminary study relating the flatness of the loss landscape of the policy network to it's apparent robustness, further research is required to quantify its susceptibility to noise.

\section{Conclusions}
Our evaluations reveal the fragile nature of vanilla RL agents trained for C2 when deployed in insecure environments where even minute perturbations to the input introduced by a malicious actor are sufficient to introduce a large variability in the agent's prediction. We analyze this susceptibility of a C2 agent when commanding the BlueForce in two custom scenarios and discuss the reasons behind such behavior and the implications from a strategic perspective. Finally, we emphasize the need to develop robust training algorithms for RL which would be critical for reliable mission planning on battlefields of the future.

\begin{dci}
The Authors declare that there is no conflict of interest.
\end{dci}

\begin{funding}
This material is based in part upon work supported by the Army Research Lab~(ARL) under Contract number W911NF-2020-221. Any opinions, findings, and conclusions or recommendations expressed in this material are those of the authors and do not necessarily reflect the views of the sponsors.    
\end{funding}

\begin{biogs}

\noindent
{\sagesf\bfseries Ahaan Dabholkar} is a PhD student in the Department of Computer Science at Purdue University working in the area of Machine Learning security. His work focuses on formulating attacks on existing ML applications and protocols and  developing robust defenses against such attacks.\\

\noindent
{\sagesf\bfseries James Zachary Hare} is currently an Electronics Engineer in the Context Aware Processing Branch at the DEVCOM Army Research Laboratory (ARL), Adelphi, MD, USA. He received the B.S., M.S., and Ph.D. degrees all in electrical engineering from the University of Connecticut, Storrs, CT, USA, in 2012, 2016, and 2018, respectively. He was a Postdoctoral Fellow at ARL from 2018-2021. His research interests include distributed learning, uncertainty analysis and modeling for AI/ML, change point detection, reinforcement learning, sensor networks, information fusion, and intelligent network control systems.\\

\noindent
{\sagesf\bfseries Mark Mittrick} is a computer scientist with the DEVCOM Army Research Laboratory located at Aberdeen Proving Ground since 2003. Prior to that, he received his B.S in Computer Science from Wilkes University in 2003 and later while working at the Proving Ground earned his M.S from Towson University in 2009. He has authored and co-authored several technical reports in the computer science field.\\

\noindent
{\sagesf\bfseries John Richardson} is a computer scientist at the U.S. Army DEVCOM Army Research Laboratory working in the Content Understanding Branch.\\

\noindent
{\sagesf\bfseries Nicholas Waytowich} is a distinguished Machine Learning Research Scientist at the U.S. Army Research Laboratory. In his capacity as the science lead for the ARL’s Human-Guided Machine Learning Branch, Dr. Waytowich spearheads innovative research in human-guided AI/ML. His work focuses on developing algorithms that integrate human feedback, utilizing human-in-the-loop machine learning and reinforcement learning techniques. His research interests are broad and include deep reinforcement learning, human-agent teaming, and human-robot collaboration. Recently, his research direction has expanded to include the use of generative AI and large language models in both defense and civilian applications.\\

\noindent
{\sagesf\bfseries Priya Narayanan} is the Chief of Autonomous Systems Branch within the Science of Intelligent Systems Division (SISD) at ARL. She provides technical leadership over basic and applied research in the areas of robotics and autonomous systems to advance the science of autonomy, robotics, and multi-agent teaming for air, ground, and unique mobility and whole-body manipulation platforms. Prior to her current role, she was the Lead of the Intelligent Situation Awareness Team within the Intelligent Perception Branch at ARL, where she provided technical guidance for research in AI/ML-based scene understanding for autonomous systems, battlefield modeling for decision support, and aligning research activities of the team to ARL’s larger portfolio of autonomy and decision dominance. She has served as the Program Manager for “AI for C2 of MDO” Director’s Strategic Initiative (DSI) program that focused on development of AI-enabled systems for Command and Control of Multi-Domain Forces.\\

\noindent
{\sagesf\bfseries Saurabh Bagchi} is a Professor of Electrical and Computer Engineering and Computer Science at Purdue University working in the area of dependable and secure software systems. He serves on the Board of the IEEE Computer Society and is the Co-Founder and CTO of a cloud computing startup, KeyByte. He is the founding Director of a university-wide resilience center at Purdue called CRISP and Director of the Army's Artificial Intelligence Innovation Institute (A2I2). Saurabh is proudest of the 25 PhD students and 30 Masters thesis students who have graduated from his research group and who are in various stages of building wonderful careers in industry or academia.

\end{biogs}


\begin{thebibliography}{10}
\providecommand{\url}[1]{\texttt{#1}}
\providecommand{\urlprefix}{URL }
\expandafter\ifx\csname urlstyle\endcsname\relax
  \providecommand{\doi}[1]{DOI:\discretionary{}{}{}#1}\else
  \providecommand{\doi}{DOI:\discretionary{}{}{}\begingroup \urlstyle{rm}\Url}\fi
\providecommand{\eprint}[2][]{\url{#2}}

\bibitem{starcraft}
Blizzard.
\newblock Starcraft II.
\newblock \urlprefix\url{https://starcraft2.blizzard.com}.

\bibitem{dota}
Valve.
\newblock Dota 2.
\newblock \urlprefix\url{https://www.dota2.com/home}.

\bibitem{vinyals}
Vinyals O, Babuschkin I, Czarnecki WM et~al.
\newblock {Grandmaster level in StarCraft II using multi-agent Reinforcement Learning}.
\newblock \emph{Nature} 2019; 575(7782): 350--354.
\newblock \doi{10.1038/s41586-019-1724-z}.

\bibitem{openai2019dota}
OpenAI, :, Berner C et~al.
\newblock Dota 2 with large scale Deep Reinforcement Learning, 2019.
\newblock \eprint{1912.06680}.

\bibitem{pysc2}
Google Deepmind.
\newblock pysc2.
\newblock \urlprefix\url{https://github.com/google-deepmind/pysc2}.

\bibitem{samvelyan2019starcraft}
Samvelyan M, Rashid T, de~Witt CS et~al.
\newblock The Starcraft Multi-Agent Challenge, 2019.
\newblock \eprint{1902.04043}.

\bibitem{smacgithub}
oxwhirl.
\newblock smacv2.
\newblock \urlprefix\url{https://github.com/oxwhirl/smacv2}.

\bibitem{pydota2}
pydota2.
\newblock pydota2.
\newblock \urlprefix\url{https://github.com/pydota2/pydota2}.

\bibitem{narayanan2021first}
Narayanan P, Vindiola M, Park S et~al.
\newblock First-year report of ARL Directors Strategic Initiative (FY20-23): Artificial Intelligence (AI) for Command and Control (C2) of Multi-Domain Operations (MDO).
\newblock \emph{US Army Combat Capabilities Development Command, Army Research Laboratory} 2021; .

\bibitem{park2022deep}
Park SJ, Vindiola MM, Logie AC et~al.
\newblock Deep Reinforcement Learning to assist Command and Control.
\newblock In \emph{Artificial Intelligence and Machine Learning for Multi-Domain Operations Applications IV}, volume 12113. SPIE, pp. 430--438.

\bibitem{soleyman2020multi}
Soleyman S and Khosla D.
\newblock Multi-agent Mission Planning with Reinforcement Learning.
\newblock In \emph{AAAI Symposium on the 2nd Workshop on Deep Models and Artificial Intelligence for Defense Applications: Potentials, Theories, Practices, Tools, and Risks}. AAAI, pp. 51--57.

\bibitem{zhang2020air}
Zhang L, Xu J, Gold D et~al.
\newblock Air Dominance through Machine Learning.
\newblock \emph{Santa Monica, CA: RAND Corporation} 2020; .

\bibitem{BZCFHSWA2022}
Basak A, Zaroukian EG, Corder K et~al.
\newblock Utility of Doctrine with multi-agent RL for Military Engagements.
\newblock In \emph{Artificial Intelligence and Machine Learning for Multi-Domain Operations Applications IV}, volume 12113. SPIE, pp. 609--628.

\bibitem{waytowich2022learning}
Waytowich N, Hare J, Goecks VG et~al.
\newblock Learning to guide multiple heterogeneous actors from a single human demonstration via Automatic Curriculum Learning in Starcraft II.
\newblock In \emph{Artificial Intelligence and Machine Learning for Multi-Domain Operations Applications IV}, volume 12113. SPIE.

\bibitem{vinyals2017starcraft}
Vinyals O, Ewalds T, Bartunov S et~al.
\newblock Starcraft II: A new challenge for Reinforcement Learning.
\newblock \emph{arXiv preprint arXiv:170804782} 2017; .

\bibitem{marr2001military}
Marr JJ.
\newblock \emph{The Military Decision Making Process: Making better decisions versus making decisions better}.
\newblock School of Advanced Military Studies, US Army Command and General Staff College, 2001.

\bibitem{s2000}
Shoffner WA.
\newblock \emph{The Military Decision Making Process: Time for a Change}.
\newblock School of Advanced Military Studies, US Army Command and General Staff College, 2000.

\bibitem{goecks2023games}
Goecks VG, Waytowich N, Asher DE et~al.
\newblock On games and simulators as a platform for development of Artificial Intelligence for Command and Control.
\newblock \emph{The Journal of Defense Modeling and Simulation} 2023; 20(4): 495--508.

\bibitem{liang2018rllib}
Liang E, Liaw R, Nishihara R et~al.
\newblock Rllib: Abstractions for Distributed Reinforcement Learning.
\newblock In \emph{International conference on machine learning}. PMLR, pp. 3053--3062.

\bibitem{mnih2016asynchronous}
Mnih V, Badia AP, Mirza M et~al.
\newblock Asynchronous methods for Deep Reinforcement Learning.
\newblock In \emph{International conference on machine learning}. PMLR, pp. 1928--1937.

\bibitem{schulman2017proximal}
Schulman J, Wolski F, Dhariwal P et~al.
\newblock Proximal Policy Optimization algorithms.
\newblock \emph{arXiv preprint arXiv:170706347} 2017; .

\bibitem{schulman2015trust}
Schulman J, Levine S, Abbeel P et~al.
\newblock Trust Region Policy Optimization.
\newblock In \emph{International conference on machine learning}. PMLR, pp. 1889--1897.

\bibitem{huang2017adversarial}
Huang S, Papernot N, Goodfellow I et~al.
\newblock Adversarial Attacks on Neural Network Policies.
\newblock \emph{arXiv preprint arXiv:170202284} 2017; .

\bibitem{sun2020stealthy}
Sun J, Zhang T, Xie X et~al.
\newblock Stealthy and efficient Adversarial Attacks against Deep Reinforcement Learning.
\newblock In \emph{Proceedings of the AAAI Conference on Artificial Intelligence}, volume~34. pp. 5883--5891.

\bibitem{wu2021adversarial}
Wu X, Guo W, Wei H et~al.
\newblock Adversarial Policy Training against Deep Reinforcement Learning.
\newblock In \emph{30th USENIX Security Symposium (USENIX Security 21)}. pp. 1883--1900.

\bibitem{gleave2019adversarial}
Gleave A, Dennis M, Wild C et~al.
\newblock Adversarial Policies: Attacking Deep Reinforcement Learning.
\newblock \emph{arXiv preprint arXiv:190510615} 2019; .

\bibitem{7958570}
Carlini N and Wagner D.
\newblock Towards Evaluating the Robustness of Neural Networks.
\newblock In \emph{2017 IEEE Symposium on Security and Privacy (SP)}. pp. 39--57.
\newblock \doi{10.1109/SP.2017.49}.

\bibitem{goodfellow2014explaining}
Goodfellow IJ, Shlens J and Szegedy C.
\newblock Explaining and Harnessing Adversarial Examples.
\newblock \emph{arXiv preprint arXiv:14126572} 2014; .

\bibitem{madry2018towards}
Madry A, Makelov A, Schmidt L et~al.
\newblock Towards Deep Learning Models Resistant to Adversarial Attacks.
\newblock In \emph{International Conference on Learning Representations}.
\newblock \urlprefix\url{https://openreview.net/forum?id=rJzIBfZAb}.

\end{thebibliography}
\end{document}